\documentclass[fleqn,usenatbib,useAMS,twocolumn,a4paper, aps]{revtex4}
\usepackage{aas_macros}
\usepackage{mathtools}
\usepackage[T1]{fontenc}
\usepackage{ae,aecompl}
 \usepackage{hyperref}
\usepackage{xcolor}

\usepackage{graphicx}	
\usepackage{amsmath}	
\usepackage{amssymb}	
\usepackage{bm}		

\begin{document}

\title [Thermal conduction in shocked accretion flow]{Role of thermal conduction in relativistic hot accretion flow around rotating black hole with shock}

\author{Monu Singh}
\author{Santabrata Das}
\email{sbdas@iitg.ac.in}
\email{monu18@iitg.ac.in}
\affiliation{Department of Physics, Indian Institute of Technology Guwahati, Guwahati, 781039, Assam, India.}

\begin{abstract}
    We investigate the properties of low angular momentum, relativistic, viscous, advective accretion flows around rotating black holes that include shock waves in the presence of thermal conduction. We self-consistently solve the governing fluid equations to obtain the global transonic accretion solutions for a set of model parameters, namely energy ($\mathcal{E}$), angular momentum ($\lambda$), viscosity ($\alpha$), conduction parameter ($\Phi_{\rm s}$) and cooling parameter ($f_{\rm c}$). We observe that depending on the model parameters, accretion flow experiences centrifugally supported shock transition and the present study, for the first time, focuses on examining the shock properties, such as shock radius ($r_{\rm s}$), compression ratio ($R$) and shock strength ($\Psi$) regulated by the dissipation parameters ($\Phi_{\rm s},~f_{\rm c}$). We show that shock-induced global accretion solutions persist for wide range of model parameters and identify the boundary of the parameter space in energy-angular momentum plane that admits standing shocks for different dissipation parameters ($\Phi_{\rm s},~f_{\rm c}$). Finally, we compute the critical conduction parameter ($\Phi_{\rm s}^{\rm cri}$), beyond which shock ceases to exist. We find that $\Phi_{\rm s}^{\rm cri}$ directly depends on the black hole spin ($a_{\rm k}$) with $\Phi_{\rm s}^{\rm cri} \sim 0.029$ and $\sim 0.04$ for weakly ($a_{\rm k}\rightarrow 0$) and rapidly ($a_{\rm k}\rightarrow 1$) rotating black hole. Furthermore, we observe that $\Phi_{\rm s}^{\rm cri}$ decreases with increasing viscosity ($\alpha$), and shocked accretion solutions continue to exist for $\alpha \lesssim 0.065$ ($a_{\rm k} \rightarrow 0$) and $\lesssim 0.104$ ($a_{\rm k} \rightarrow 1$), respectively.
\end{abstract}

 
	

\maketitle

\section{Introduction}
\label{intro}

Accretion onto black holes is indeed an appealing process with profound astrophysical implications as it successfully manifests the energetic electromagnetic radiations emergent from X-ray binaries (XRBs) and active galactic nuclei (AGNs) \cite[]{Frank-etal2002}. Over the past several decades, numerous efforts have been made to develop theoretical models aimed in understanding the underlying physical processes in accretion flow. The seminal works of \cite{shakura-sunyaev-1973,Novikov-Thorne1973} played major role in developing a geometrically thin and optically thick accretion disk that laid the foundation for understanding the observed characteristics of accreting systems. \cite{Abramowicz-etal1988} proposed a model describing the optically thick, geometrically slim accretion disks, where advective cooling prevails over radiative cooling at high accretion rates. Later on, a model of advection-dominated accretion flow \cite[ADAF;][]{Narayan-Yi1994, Narayan-Yi1995} featuring a hot, geometrically thick, optically thin disk at low accretion rate (much lower than Eddington limit) fascinated the astrophysicists because of its tremendous success in studying the black hole sources across the wide mass range including both XRBs \cite[and references therein]{Esin-etal1997,Hameury-etal1997,Yuan-Cui2005,Liu-etal2011} and AGNs \cite[and references therein]{Reynolds-etal1996,Manmoto-etal1997,Yuan-Narayan2014,Younes-etal2019}.

Indeed, when the central source accretes matter at very low rate, the temperature and density profiles of the hot accreting matter resemble the radiatively inefficient accretion flow (RIAF) due to its reduced radiative efficiency \cite[]{Ichimaru1977,Yuan-Narayan2014}. In this scenario, the accreting plasma is expected to remain in the weakly collisional regime with mean free paths exceeding the typical length scale of the accretion disk structure, equivalently the gravitational radius $r_{\mathrm{g}} = GM_{\rm BH}/c^2$, where $G$, $M_{\rm BH}$, and $c$ denote the gravitational constant, the BH mass, and speed of light, respectively \cite[]{tanaka-menou-2006,Johnson-Quataert2007}. Because of this, thermal conduction becomes dynamically important, allowing the accreting matter to transport energy via heat flux, and this is the focus of the present paper.

Meanwhile, efforts have been made to examine the effect of thermal conduction on accretion solutions adopting the self-similar formalism while investigating outflows \cite[][]{tanaka-menou-2006,shadmehri-2008,Faghei2012,Khajenabi-Shadmehri-2013,ghoreyshi-shadmehri-2020}. In particular, \cite{tanaka-menou-2006} indicated that thermal conduction possibly seems to play a decisive role in launching bipolar rotating outflows from radiatively inefficient hot accretion flows. \cite{shadmehri-2008} reported that thermal conduction serves as an additional heating source that possibly redirects a part of the inflowing matter as outflows. Furthermore, \cite{Mosallanezhad-etal2021} performed numerical simulations of hot accretion flows including thermal conduction and illustrates its impact on modifying the structure of the accretion flow with a steeper density profile. \cite{Bu-etal2016} carried out related numerical analysis and claim that thermal conduction can enhance wind velocity, resulting in an approximately tenfold increase in the wind energy flux. Very recently, \cite{mitra-etal-2023} investigated global accretion solutions and demonstrated that the presence of thermal conduction significantly alters their transonic properties. All of these studies collectively emphasize the pivotal role of thermal conduction in comprehending the structure of accretion flows around black holes.

It is worth mentioning that during the accretion process, infalling matter undergoes a sonic state transition from subsonic to supersonic as it crosses a critical point before entering the black hole \cite[and references therein]{Mitra-Das2024}. Additionally, rotating matter encounters centrifugal repulsion while accreting towards the black hole, leading to the accumulation of infalling material in its vicinity. However, the accumulation of matter cannot continue indefinitely; beyond a certain threshold, the centrifugal barrier triggers a discontinuous shock transition in the flow variables \cite[and references therein]{Fukue1987,Chakrabarti1989,chakrabarti-1996,Das-etal2001,das-2007}. It is important to note that global accretion solutions harboring shock waves are thermodynamically preferred due to their higher entropy content compared to shock-free solutions \cite[]{Becker-Kazanas2001}. Meanwhile, the issue of shock formation in accretion disks around black holes has been investigated by various authors both theoretically \cite[]{Fukue1987,Chakrabarti1989,Yang-Kafatos1995,chakrabarti-1996,lu-etal-1999,Das-etal2001,Becker-Kazanas2001,chakrabarti-das-2004,Le-Becker2004,Fukumura-Tsuruta2004,das-2007,Becker-etal2008,Das-etal2009,Kumar-etal2013,Sarkar-Das2016,Aktar-etal2017,dihingia-etal-2019A,Sen-etal2022,Singh-Das2024,Mitra-Das2024} and by means of numerical simulations \cite[]{Chakrabarti-Molteni1993,Molteni-etal1994,Molteni-etal1996,Ryu-etal1997,Lanzafame-etal1998,Das-etal2014,Okuda-Das2015,Sukova-Janiuk2015,Lee-atal2016,Sukova-etal2017,Kim-etal2019,Okuda-etal2019,Palit-etal2019,Debnath-etal2024,TianLe-Zhao-etal2024}. All these works draw attention as shock-induced global accretion solution satisfactorily explains the spectro-temporal properties of black hole candidates \cite[]{chakrabarti-titarchuk-1995,nandi-etal-2012,Iyer-etal2015,Nandi-etal2018,Das-etal2021,Majumder-etal2023}. However, the investigation of shocked accretion solutions in the presence of thermal conduction has not yet been explored in the astrophysical literature.

Being motivated with this, in this work, we examine the effect of the thermal conduction on the shock-induced global accretion solutions around rotating black hole. These solutions are especially pertinent for describing the dynamics of weakly collisional flows in black hole systems that accrete matter at extremely low rates, such as those observed in Sgr A* and M87 \cite[]{EHT-M87-2021,EHT-SGRA*-2022}. In these systems, thermal conduction becomes important resulting in a transport of energy flux from the inner to the outer regions of the accretion disk \cite[]{Johnson-Quataert2007,Quataert2008}. Since thermal conduction predominantly affects the outer regions of the disk \cite[$r\gtrsim{\rm few~tens~of~gravitational~radius~r_g}$;][]{mitra-etal-2023}, it is highly likely that thermal conduction exerts significant influence on global accretion solutions harboring shock waves. Considering this, we study the properties of standing shock, namely shock radius ($r_{\rm s}$), compression ratio ($R$) and shock strength ($\Psi$) in terms of the model parameters and find that shock continues to form even in presence of high thermal conduction around both weakly ($a_{\rm k} \rightarrow 0$) and rapidly ($a_{\rm k} \rightarrow 1$) rotating black holes. We further identify the parameter space, spanned by the energy and angular momentum of the flow, that permits shocked accretion solutions, and investigate how this parameter space is altered by the influence of thermal conduction. Finally, we determine the critical value of thermal conduction ($\Phi^{\rm cri}_{\rm s}$) that admits standing shocks in dissipative accretion flow and observe that $\Phi^{\rm cri}_{\rm s}$ exhibits strong dependencies on $\alpha$, $f_{\rm c}$ and $a_{\rm k}$.

The main structure of the paper is as follows. In Section 2, we present the model assumption and the governing equations that describe the flow motion around black hole. We discuss the obtained results involving shocked accretion solutions, shock properties, shock parameter space and critical conduction parameter for standing shocks in Section 3. Finally, we summarize our findings in Section Section 4.

\section{GOVERNING EQUATIONS}

We consider low angular momentum, axisymmetric, viscous, advective accretion flow around a rotating black hole. In order to describe the accretion flow, we use cylindrical coordinate system ($r,\phi,z$) with the black hole resides at the origin. In addition, we consider that the accretion flow maintains hydrostatic equilibrium in the vertical direction (along $z$-direction) and therefore, flow variables are vertically integrated. Furthermore, we choose $M_{\rm BH}=G=c=1$, where $M_{\rm BH}$ is the black hole mass, $G$ is the gravitational constant and $c$ refers the speed of light. In this unit system, radial distance ($r$), specific angular momentum ($\lambda$) and specific energy ($\mathcal{E}$) are expressed in units of $r_{\rm g}=GM_{\rm BH}/c^2$, $GM_{\rm BH}/c$, and $c^2$, respectively. 

The spacetime geometry of a rotating Kerr-like black hole is satisfactorily described by an effective potential ($\Phi^{\textrm{eff}}$) \cite[]{dihingia-etal-2018} given by,
\begin{equation}
\label{eqn:phi}
  \Phi_{\textrm{e}} ^{\textrm{eff}} = \frac{1}{2}\ln\left[\frac{r \Delta}{a_{\textrm{k}}^2 (r+2) - 4 a_{\textrm{k}} \lambda + r^3 -\lambda^2(r-2)}\right],  
\end{equation}
where $r$ denotes the radial coordinate, $a_{\rm k}$ is the spin of the black hole, and $\Delta = r^2 - 2 r + a_{\textrm{k}}^2$. The suffix `e' refers the quantity measured at the disk equatorial plane.

The governing hydrodynamic equations that describe the axisymmetric accretion flow in the steady state are given by:

\noindent (a) Mass flux conservation equation:
\begin{equation}
\label{eqn:continuity}
	\dot{M} = 2\pi \upsilon \Sigma \sqrt{\Delta}\thickspace,
\end{equation}
\noindent (b) The radial momentum equation: 
\begin{equation}
\label{eqn:momentum_r}
\upsilon \frac{d\upsilon}{dr}+\frac{1}{h \rho}\frac{dP}{dr}+\frac{d \Phi_{\textrm{e}}^{\textrm{eff}}}{dr} = 0,
\end{equation}
\noindent (b) The azimuthal momentum equation:
\begin{equation}
\label{eqn:momentum_phi}
	\upsilon \frac{d\lambda}{dr} + \frac{1}{\Sigma\space r}\frac{d}{dr}(r^2 W_{r\phi}) = 0,
\end{equation}
and
\noindent (d) The energy equation:

\begin{equation}
\label{eqn:energy}
\frac{\upsilon}{\Gamma - 1}\left(\frac{dP}{dr} - \frac{\Gamma P}{\rho}\frac{d\rho}{dr}\right) = f_{\rm c}\alpha r(P+ \rho \upsilon^2) \frac{d \Omega}{d r} + \frac{1}{r} \frac{d (r F_{s})}{d r}
\end{equation}
The local variables $\upsilon$, $\rho$, $P$ and $\Sigma$ denote the radial velocity, mass density, isotropic pressure, and vertically integrated density of the flow. In equation (\ref{eqn:momentum_phi}), $W_{r\phi}$ refers the vertically integrated viscous stress and is given by 
\begin{equation}
    W_{r\phi}= -\alpha\Pi= - \alpha \left(W + \Sigma \upsilon^2 \right),
    \label{eqn:vis-strees}
\end{equation}
where $\alpha$ is the viscosity parameter that determines the angular momentum transport inside the disk, and $W$ and $\Pi~(= W + \Sigma \upsilon^2)$ are the vertically integrated gas pressure and total pressure, respectively. In equation (\ref{eqn:energy}), $\Gamma$ is the adiabatic index, and $H$ is the local half-thickness of the disk, which is expressed as \cite[]{Riffert-Herold1995,Peitz-Appl1997}, 
$$
H^2 = \frac{P r^{3}}{\rho \mathcal{F}}, \thickspace \mathcal{F} = \frac{1}{(1-\lambda \Omega)}\times  \frac{(r^2 + a_{\textrm{k}}^{2})^{2} + 2 \Delta a_{\textrm{k}}^{2}}{(r^2 + a_{\textrm{k}}^{2})^{2} - 2 \Delta a_{\textrm{k}}^{2}},
$$
where $\Omega$ is angular velocity of the flow and given by,
$$
\Omega = \frac{2 a_{\textrm{k}} + \lambda (r - 2)}{a_{\textrm{k}}^{2}(r + 2) - 2 a_{\textrm{k}}\lambda + r^3} \thickspace.
$$
In the first term on the right-hand side of equation (\ref{eqn:energy}), we introduce the parametric cooling factor $f_{\rm c} ~(=1-Q_{\rm rad}/Q_{\rm vis})$, where $Q_{\rm vis}$ and $Q_{\rm rad}$ represent the viscous heating and radiative cooling rates \cite[]{Narayan-Yi1994, mitra-etal-2023}. In this study, we assume $f_{\rm c}$ to be constant throughout the flow, with $0 \le f_{\rm c} \le 1$. When the flow is advection-dominated, $f_{\rm c}=1$. In contrast, $f_{\rm c}=0$ indicates a flow dominated by radiative cooling. The second term on the right-hand side of equation (\ref{eqn:energy}) accounts for the energy transfer due to saturated thermal conduction \cite[]{tanaka-menou-2006}. Here, $F_{\rm s}$ denotes the saturated conduction flux \cite[]{cowie-mcKee-1977}, and is defined as $F_{\rm{s}} = 5 \Phi_{\rm{s}} \rho \left(\frac{P}{\rho}\right)^{3/2}$, where $\Phi_{\rm{s}}$ is the dimensionless saturated conduction parameter (hereafter conduction parameter) having values in the range $0 \le \Phi_{\rm s} \le 1$, and it controls the effect of thermal conduction inside the disk. Note that $\Phi_{\rm s}$ and $f_{\rm c}$ are often collectively referred to as dissipation parameters.

We close the governing equations (\ref{eqn:continuity}-\ref{eqn:energy}) using an equation of state (EoS) that relates the internal energy ($\epsilon$), pressure ($P$) and density ($\rho$) of the accreting matter. In reality, the thermally non-relativistic (adiabatic index $\Gamma=5/3$) flow at the outer edge becomes thermally relativistic ($\Gamma=4/3$) upon reaching the inner edge of the disk. Hence, to deal with thermally relativistic flow, we choose a relativistic EoS in accommodating the variable adiabatic index ($\Gamma$). For accreting fluid consisting of ions and electrons, the relativistic EoS is given by \cite[]{chattopadhyay-ryu-2009},
\begin{equation}
    \epsilon = n_{\textrm{e}} m_{\textrm{e}} \rho = \frac{\rho}{\tau} F, \quad {\rm and} \quad P=\frac{2\rho\Theta}{\tau},
    \label{eqn:eos}
\end{equation}
where $\tau = 1+ m_p/m_e$ and 
$$
F=\left[ 1+ \Theta \left( \frac{9\Theta +3}{3 \Theta +2}\right) \right] + \left[ \frac{m_p}{m_e} + \Theta \left( \frac{9\Theta m_e +3m_p}{3 \Theta m_e+ 2 m_p}\right) \right],
$$
Here, $m_p$ and $m_e$ denote the masses of ion and electron, respectively, and $\Theta~(=k_{\rm B}T/m_e c^2)$ is the dimensionless temperature, $T$ is the temperature in Kelvin and $k_{\rm{B}}$ is the Boltzmann constant. We define sound speed as $C_s = \sqrt{2\Gamma\Theta/(F+2\Theta)}$, where $\Gamma ~ [= (1+N)/N]$ refers the adiabatic index and $N ~[=(1/2)(dF/d\Theta)]$ is the polytropic index of the flow, respectively \cite[]{dihingia-etal-2019A}.

Following \cite{chattopadhyay-ryu-2009} and using equation (\ref{eqn:continuity}), we calculate the entropy accretion rate as,
$$
\dot{\mathcal{M}} = \upsilon H \sqrt{\Delta} \left[ \Theta^{2} (2+3\Theta)(3\Theta +2 m_{p}/m_{e}) \right]^{3/2} \exp{(k_{1})},
$$ 
where $k_{1} = \left[F- (1+m_{p}/m_{e}) \right]/2\Theta$.

We simplify equations (\ref{eqn:continuity}), (\ref{eqn:momentum_r}), (\ref{eqn:momentum_phi}), (\ref{eqn:energy}) and (\ref{eqn:eos}) and obtain the coupled differential equations involving flow variables as
\begin{equation}
    R_{0}+ R_{\upsilon}\frac{d \upsilon}{d r} + R_{\Theta}\frac{d \Theta}{d r} + R_{\lambda}\frac{d \lambda}{d r} = 0
    \label{eqn:Rdr}
\end{equation}
\begin{equation}
    L_{0} + L_{\upsilon}\frac{d \upsilon}{d r} + L_{\Theta}\frac{d \Theta}{d r} + L_{\lambda}\frac{d \lambda}{d r} = 0 ,
    \label{eqn:Ldr}
\end{equation}
\begin{equation}
    E_{0} + E_{\upsilon}\frac{d \upsilon}{d r} + E_{\Theta}\frac{d \Theta}{d r} + E_{\lambda}\frac{d \lambda}{d r} = 0.
    \label{eqn:Edr}
\end{equation}
The explicit mathematical expression of the coefficients ($R_{i}$, $L_{i}$, $E_{i}$, $i\rightarrow 0$, $\upsilon$, $\Theta$ and $\lambda$) in equations (\ref{eqn:Rdr}), (\ref{eqn:Ldr}) and (\ref{eqn:Edr}) are intricate in nature and provided in the Appendix.

Using equations (\ref{eqn:Rdr}), (\ref{eqn:Ldr}) and (\ref{eqn:Edr}), we obtain the wind equation given by, 
\begin{equation}
    \frac{d\upsilon}{d r} = \frac{\mathcal{N}(r,\upsilon,\lambda,\Theta)}{\mathcal{D}(r,\upsilon,\lambda,\Theta)},
    \label{eqn:nbyd}
\end{equation}
where the numerator $\mathcal{N}(r,\upsilon,\lambda,\Theta)$ and denominator $\mathcal{D}(r,\upsilon,\lambda,\Theta)$ are functions of flow variables and their explicit expressions are provided in Appendix. In addition, we calculate the gradient of $\Theta$ and $\lambda$ as,
\begin{equation}
    \frac{d\Theta}{d r} = \Theta_{1}+\Theta_{2}\frac{d\upsilon}{d r}, \quad {\rm and} \quad \frac{d\lambda}{d r} = \lambda_{1}+\lambda_{2}\frac{d\upsilon}{d r},
    \label{eqn:dtheta}
\end{equation}
where $\Theta_1$, $\Theta_2$, $\lambda_1$ and $\lambda_2$ are described in Appendix.

We self-consistently solve the coupled differential equations (\ref{eqn:nbyd}) and (\ref{eqn:dtheta}) for a set of model parameters to obtain the global transonic accretion solution around black hole. In doing so, we treat $\alpha$, $\Phi_{\rm s}$ and $a_{\rm k}$ as global parameters. Further, we set a reference radius $r_{\rm ref}$ and the angular momentum ($\lambda_{\rm ref}$) at $r_{\rm ref}$ as local parameters to initiate the numerical integration of these equations. As the black hole solutions are inherently transonic, the flow inevitably passes through the critical point. Hence, following the approach outlined in \cite{chakrabarti-das-2004}, we choose critical point ($r_{\rm c}$) as the reference radius $i.e.$, $r_{\rm c}=r_{\rm ref}$ and integrate equations (\ref{eqn:nbyd}) and (\ref{eqn:dtheta}) from $r_{\rm c}$ inward up to the horizon and then outward to a distant radius (hereafter `disk outer edge $r_{\rm edge}$'). Subsequently, we combine these two parts of the solution to get a comprehensive global transonic accretion solution. It is worth mentioning that depending on the local and global parameters, accretion flow may contain either single or multiple critical points \cite[and references therein]{das-2007,das-etal-2021}. A critical point that forms close to the horizon is referred to as the inner critical point ($r_{\rm in}$), whereas a critical point situated further from the horizon is referred to as the outer critical point ($r_{\rm out}$). It is worth mentioning that the presence of multiple critical points is a prerequisite for triggering a shock transition in an accretion flow around black hole \cite{Das-etal2001}.

\section{RESULTS}

\subsection{Global Transonic Accretion Solutions}
\label{sec:GTAS}

\begin{figure}
	\begin{center}
		\includegraphics[width=\columnwidth]{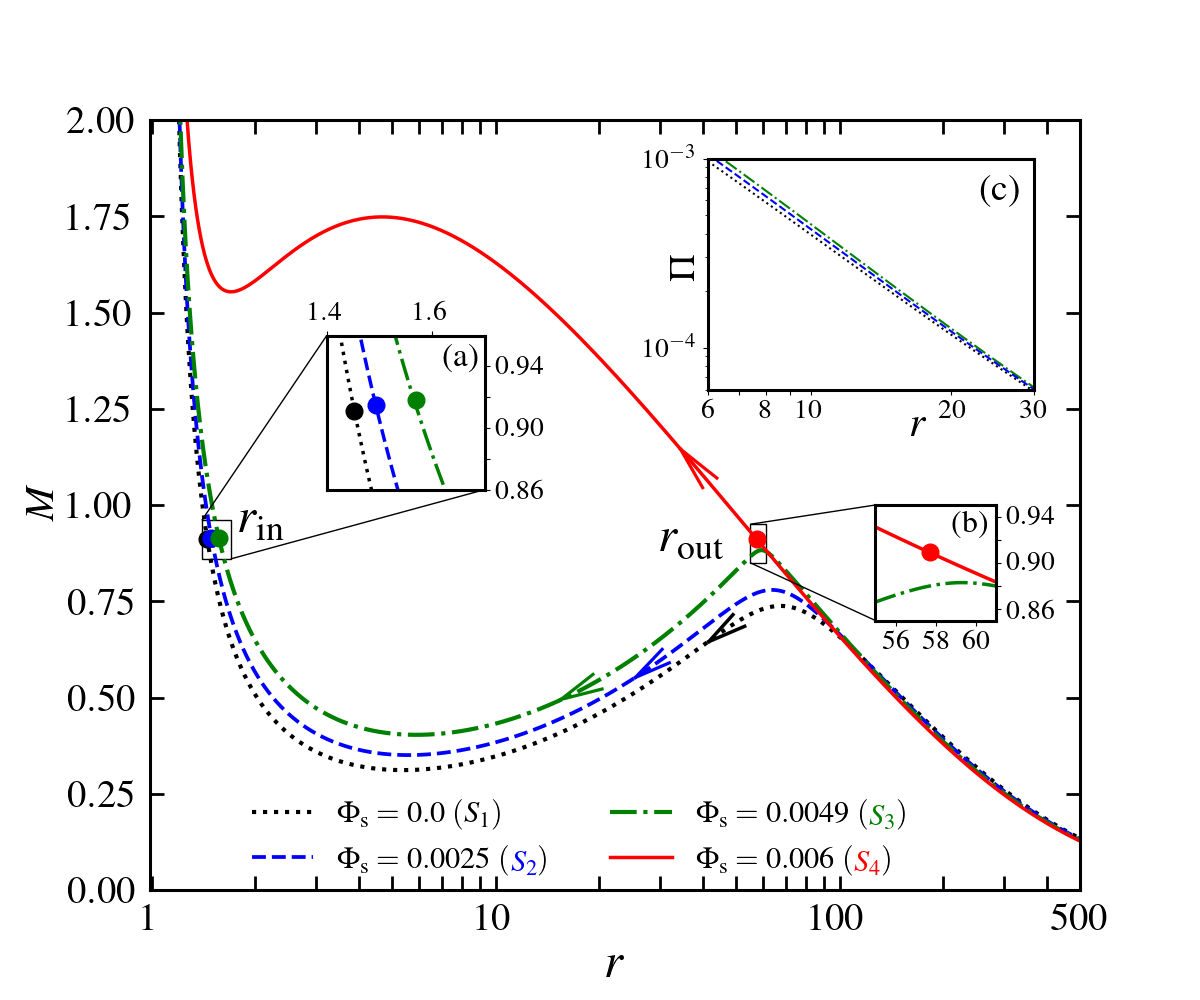}
	\end{center}
	\caption{Plot of Mach number ($M=\upsilon/C_{\rm s}$) of the accreting flow as function of radial coordinate ($r$) for different conduction parameters ($\Phi_{\rm s}$). Flows are injected from the outer edge of the disk $r_{\rm{edge}} = 500$ with $\lambda_{\rm{edge}} = 3.294$, $\mathcal{E_{\rm{edge}}}=1.0036$, and $\alpha=0.01$ on to a black hole of spin $a_{\rm k}=0.99$. Flow solutions depicted using dotted (black), dashed (blue), and dot-dashed (green) are obtained for $\Phi_{\rm s} = 0.0~(S_1)$, $0.0025~(S_2)$ and $0.0049~(S_3)$, while solid (red) curve denotes solution corresponding to $\Phi_{\rm s}= 0.006~(S_4)$. Critical points are marked ($r_{\rm in}$ and $r_{\rm out}$) in the figure. Insets (a) and (b) zoom the locations of the critical points for clarity, and inset (c) illustrates the pressure profiles. See the text for details.   
	}
	\label{fig:MrPhis}
\end{figure}

\begin{table*}
    \caption{Details of the flow variables corresponding to the global transonic solutions presented in Fig. \ref{fig:MrPhis}. In column $2-10$, conduction parameter ($\Phi_{\rm s}$), critical point location ($r_{\rm c}$), angular momentum at $r_{\rm c}$ ($\lambda_{\rm{c}}$), velocity at $r_{\rm c}$ ($\upsilon_{c}$), temperature at $r_{\rm c}$ ($\Theta_{\rm{c}}$), outer edge of the disk ($r_{\rm edge}$), angular momentum at $r_{\rm edge}$ ($\lambda_{\rm edge}$), velocity at $r_{\rm edge}$ ($\upsilon_{\rm edge}$), temperature at $r_{\rm edge}$ ($\Theta_{\rm edge}$) are presented. See the text for more details.}
    \label{tab:table-1}
    \begin{tabular}{cc c c c c c c c c} \hline \hline
    Solution&$\Phi_{\rm{s}}$&$r_{\textrm{c}}$&$\lambda_{\textrm{c}}$&$\upsilon_{\textrm{c}}$&$\Theta_{\textrm{c}}$&$r_{\textrm{edge}}$&$\lambda_{\textrm{edge}}$&$\upsilon_{\textrm{edge}}$&$\Theta_{\textrm{edge}}$ \\
            &&($r_{\textrm{g}}$)&($r_{\textrm{g}} c$)&$c$&($m_{\textrm{e}} c^{2}/k_{\rm{B}} $)&($r_{\textrm{g}}$)&($r_{\textrm{g}} c$)&$c$&($m_{\textrm{e}} c^{2}/k_{\rm{B}}$)\\ 
            \hline
            $S_1$ &0.0&1.452&1.980&0.3611&161.567&500&3.294&0.006898&1.67759 \\
            $S_2$ &0.0025&1.507&1.953&0.3642&165.545&500&3.294&0.006770&1.67784 \\
            $S_3$ &0.0049&1.570&1.928&0.3647&165.993&500&3.294&0.006653&1.67807 \\
            $S_4$ &0.006&57.691&1.955&0.0816&5.244&500&3.294&0.006547&1.67827 \\
            \hline
    \end{tabular}
\end{table*}

In Fig. \ref{fig:MrPhis}, we illustrate typical examples of accretion solutions, where Mach number ($M=\upsilon/C_s$) is plotted as function of radial coordinate ($r$). The dotted (black) curve denotes accretion solution that connects the black hole horizon with the outer edge of the disk $r_{\rm edge}=500$ and it passes through the inner critical point at $r_{\rm in} = 1.452$ with angular momentum $\lambda_{\rm in}=1.980$, $\alpha=0.01$ and $\Phi_{\rm s}=0.0$ while accreting on to a black hole of spin $a_{\rm k}=0.99$. For this solution, we note the pertinent flow variables at the outer boundary as $\lambda_{\rm edge}$, $\upsilon_{\rm edge}$, $\Theta_{\rm edge}$. Interestingly, when the integration (equations (\ref{eqn:nbyd}) and (\ref{eqn:dtheta})) is carried toward the horizon using the aforementioned outer boundary values, an identical solution ($S_1$) is obtained. We calculate the local specific energy (hereafter energy) of the flow as $\mathcal{E} = (\upsilon^2/2) + \log(h) + \Phi_{\textrm{e}}^{\textrm{eff}}$ and at $r_{\rm edge}$ we have $\mathcal{E}_{\rm edge} = 1.0036$. Next, we switch on thermal conduction by setting $\Phi_{\rm s}=0.0025$ while keeping the remaining model parameters unchanged at $r_{\rm edge}$ as $\lambda_{\rm edge}=3.294$, $\mathcal{E}_{\rm edge}=1.0036$, $\alpha=0.01$, and calculate global transonic accretion solutions by suitably tuning $\upsilon_{\rm{edge}}$ and $\Theta_{\rm{edge}}$. The obtained result is plotted using dashed curve ($S_2$, blue) that passes through the inner critical point at $r_{\rm in}=1.507$. Note that $r_{\rm in}$ is shifted outward with increasing $\Phi_{\rm s}$ (see inset (a)) and these findings are in agreement with the results reported by \cite{mitra-etal-2023}. Upon increasing the saturation constant ($\Phi_{\rm s}$), we continue to obtain the similar solutions containing $r_{\rm in}$ up to the limiting value $\Phi_{\rm s} = 0.0049$ (denoted as $S_3$, dot-dashed in green). Beyond this value, the global accretion solution changes its overall character as it passes through the outer critical point ($r_{\rm out}$) instead of the inner critical point ($r_{\rm in}$). An example of such a solution is obtained for $\Phi_{\rm s} = 0.006$, which passes through $r_{\rm out} = 57.691$ (see inset (b)). This solution ($S_4$) is depicted in the figure with a solid (red) curve. Based on the above findings, we indicate that thermal conduction plays a decisive role in determining the nature of the accretion solutions, specifically whether it passes through $r_{\rm{in}}$ or $r_{\rm{out}}$. The details of the flow variables measured at the critical point ($r_{\rm in}$ or $r_{\rm out}$) and the outer edge of the disk ($r_{\rm edge}$) for these global transonic accretion solutions are presented in Table \ref{tab:table-1}. Additionally, we compute the total pressure ($\Pi$) of the accretion solutions $S_1$, $S_2$ and $S_3$, and present the results obtained in the inset (c). We find that for a fixed $r$, $\Pi$ increases with $\Phi_{\rm s}$. This happens because increasing the saturated conduction ($\Phi_{\rm s}$) leads to more heat flow from the inner hotter regions, resulting in a local increase in both temperature and gas pressure \cite[]{tanaka-menou-2006}.

\begin{figure}
    \begin{center}
        \includegraphics[width=\columnwidth]{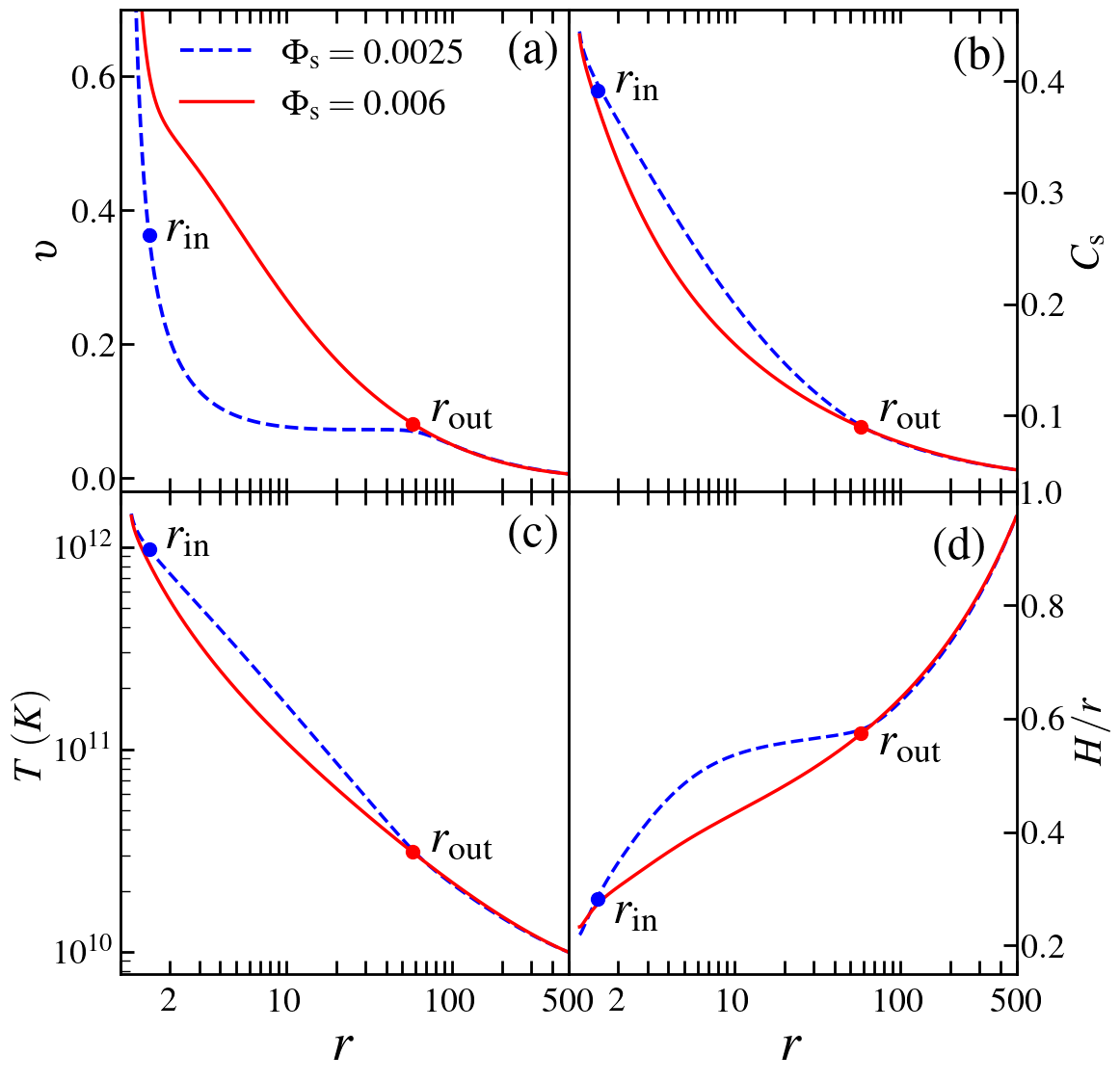}
    \end{center}
    \caption{Comparison of (a) radial velocity $\upsilon$, (b) sound speed $C_{\rm s}$, (c) temperature $T$, and (d) disk scale height $H/r$ for solutions obtained for different $\Phi_{\rm s}$. Dashed (blue) and solid (red) curves represent results for solutions labeled $S_2$ and $S_4$ in Fig. \ref{fig:MrPhis}. Filled circles denote critical points ($r_{\rm in}$ and $r_{\rm out}$). See the text for details.}
    \label{fig:flow-variables}
\end{figure}

Since the behavior of the accretion solutions passing through the inner ($r_{\rm in}$) and outer ($r_{\rm out}$) critical points is characteristically different, it is useful to examine the profiles of the flow variables for these solutions. Accordingly, we compare the radial variation of the flow variables, namely the flow radial velocity ($\upsilon$), the sound speed ($C_{\rm{s}}$), the temperature in Kelvin ($T$) and disk scale height ($H/r$), and depict the obtained results in Fig. \ref{fig:flow-variables}, where the dashed (blue) and solid (red) curves denote the results corresponding to the accretion solutions $S_2$ and $S_4$ presented in Fig. \ref{fig:MrPhis}. We observe that $\upsilon$ of $S_2$ remains lower compared to the same of $S_4$ particularly in the subsonic region ($r \lesssim 100$), despite that both solutions start with comparable radial velocity from $r_{\rm edge}=500$ as shown in panel (a). In order to preserve the conservation of mass flux (see equation (\ref{eqn:continuity})), the density ($\rho$) of the flow for $S_2$ increases, leading to an increase in both the radial profile of the sound speed ($C_{\rm s}$, panel b) and the flow temperature ($T$, panel c). This ultimately results in a higher disk-scale height $H/r$, causing a relatively thicker inner disk geometry for $S_2$.

\begin{figure}
    \begin{center}
        \includegraphics[width=\columnwidth]{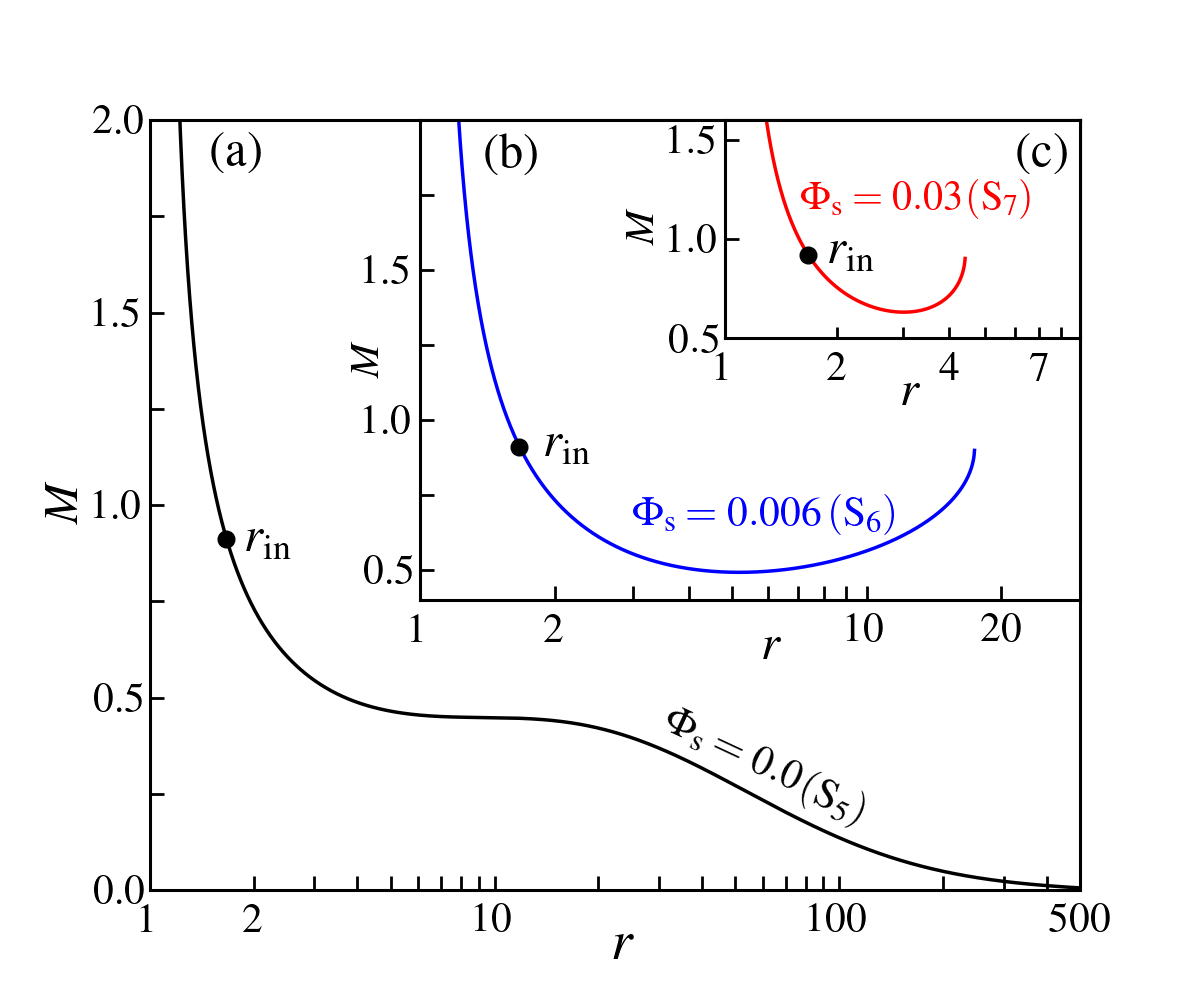}
    \end{center}
    \caption{Modification of accretion solutions ($M$ vs. $r$) passing through the inner critical point ($r_{\rm in}$) for increasing conduction parameter $\Phi_{\rm s}$. Here, we choose $r_{\rm in} = 1.667$, $\lambda_{\rm in} = 1.903$, $\alpha = 0.01$, and $a_{\rm k} = 0.99$. In panels (a), (b) and (c), results are depicted for $\Phi_{\rm s} =0.0$, $0.006$ and $0.03$, respectively. See the text for details.    
    }
    \label{fig:Mr-rin}
\end{figure}

Next, we highlight another distinct characteristic features of transonic accretion solutions around rotating black hole. In Fig. \ref{fig:Mr-rin}, we demonstrate the Mach number ($M$) profile of the accretion solutions passing through the inner critical point ($r_{\rm in}$) for different conduction parameter values $\Phi_{\rm s}$. Here, we fix the model parameters at $r_{\rm in}=1.667$ as $\lambda_{\rm in}=1.903$, $\alpha = 0.01$ and $a_{\rm k}=0.99$. In panel (a), we set $\Phi_{\rm s}=0.0$ and find that the subsonic accretion flow ($S_5$) from $r_{\rm edge}=500$ enters the black hole supersonically after crossing the inner critical point. Thereafter, we increase the thermal conduction as $\Phi_{\rm s}=0.006$ and because of that flow solution ($S_6$) becomes closed \cite[see][for details]{das-2007} for the same model parameters as shown in panel (b). This findings evidently indicate that for a set of model parameters, there exist a critical saturation constant ($\Phi^{\rm cri}_{\rm s}$) for which open solutions becomes closed. We elaborately discuss the significance of the critical saturation constant in Section \ref{sec:Cri-Phis}. It is noteworthy that accretion solution of this kind fails to connect the outer edge of the disk unless it joins with another accretion solution (similar to $S_4$ of Fig. \ref{fig:MrPhis}) via shock transition
resulting in a shock-induced global accretion solutions around black hole (see section \ref{sec:shock}). As thermal conduction increases gradually, the closed solution shrinks further, as illustrated in panel (c) for $\Phi_{\rm s} = 0.03~(S_7)$, and ultimately vanishes for $\Phi_{\rm s} > 0.11$.

\subsection{Shock-induced Global Accretion Solutions}
\label{sec:shock}

As outlined in the previous section \ref{sec:GTAS}, rotating accretion flows around black hole may harbor shock waves \cite[and references therein]{Fukue1987,Chakrabarti1989,chakrabarti-das-2004,Fukue2019,dihingia-etal-2019A,Mitra-Das2024} reminiscent of those observed in solar winds \cite[]{Holzer-Axford1970}. In a realistic scenario, subsonic accretion flow first passes through the outer critical point ($r_{\rm out}$) and continues to proceed towards the horizon supersonically. Meanwhile, rotating flow starts experiencing centrifugal repulsion that eventually triggers the discontinuous transition of the flow variables at the subsonic branch in the form of shock waves. The post-shock flow gradually picks up its velocity and eventually enters the black hole supersonically after passing through the inner critical point ($r_{\rm in}$). Indeed, the formation of standing shock is possible provided the Rankine-Hugoniot conditions (RHCs) \cite[]{Landau-Lifshitz1959} are satisfied. In this work, we employ RHCs across the shock front which are expressed as the conservation of (a) mass flux: ${\dot M}_{-}={\dot M}_{+}$, (b) energy flux : ${\mathcal E}_{-}={\mathcal E}_{+}$ and (c) momentum flux: $\Pi_{-} = \Pi_{+}$. Here, we consider the shock to be thin, and `$-$' and `$+$' refer the quantities measured across the shock front. Furthermore, in absence of any excess torque at the shock radius, we presume that the angular momentum remains continuous across the shock front, although this assumption may be violated in presence of structured magnetic fields \cite[and references therein]{Chakrabarti1989,chakrabarti-das-2004}.

\begin{figure}
    \begin{center}
        \includegraphics[width=\columnwidth]{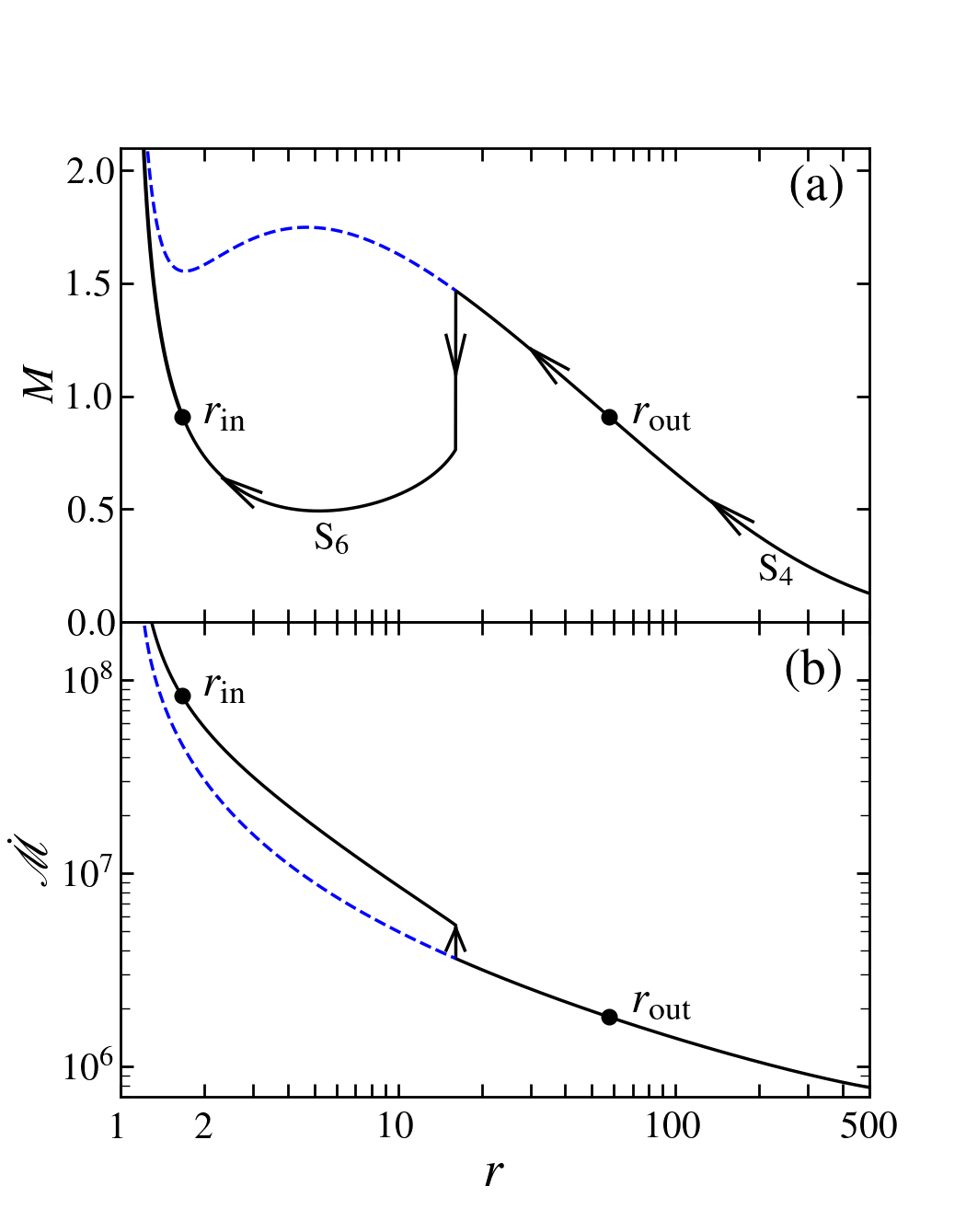}
    \end{center}
    \caption{Example of complete global accretion solution around black hole containing shock. In panel (a), the variation of Mach number ($M$) with the radial coordinate ($r$) is shown for $\Phi_{\rm s}=0.006$. Here, RHCs join $S_4$ of Fig. \ref{fig:MrPhis} and $S_{6}$ of Fig. \ref{fig:Mr-rin} at $r_{\rm s}=16.142$ (vertical arrow) to yield shock-induced global accretion solution. In panel (b), entropy accretion rate ($\dot {\mathcal{M}}$) is plotted with $r$ for the solution presented in panel (a). Filled circles denote the critical points ($r_{\rm in}$ and $r_{\rm out}$). See the text for details.}
    \label{fig:shock}
\end{figure}

In Fig. \ref{fig:shock}a, we present an example of a shock-induced global accretion solution around a rapidly rotating black hole ($a_{\rm k}=0.99$) in the presence of thermal conduction. We find that a transonic accretion solution passing through $r_{\rm out}=57.691$ ($S_4$ in Fig. \ref{fig:MrPhis}) experiences a shock transition at $r_{s}=16.142$ as the RHCs are satisfied there and jumps into the subsonic branch of another solution ($S_6$ in Fig. \ref{fig:Mr-rin}) avoiding dashed part of $S_4$ solution. After the shock, the subsonic flow gradually gains speed as it progresses inwards and ultimately plunges into the black hole supersonically after passing through the inner critical point $r_{\rm{in}} = 1.667$ (see Fig. \ref{fig:Mr-rin}b). This happens because the entropy accretion rate ($\dot{\mathcal{M}}$) in the post-shock flow is greater than the pre-shock flow, as depicted in Fig \ref{fig:shock}b, and high-entropy solutions are thermodynamically preferred \cite[]{Becker-Kazanas2001}. In the figure, the vertical arrow marks the shock transition radius, while the other arrows show the direction of the flow moving towards the black hole. Due to shock compression, the convergent post-shock flow becomes hot and dense yielding a puffed up inner disk (equivalently post-shock corona, hereafter PSC \cite[]{Aktar-etal2017}). In order to quantify the density compression across the shock front, we compute the compression ratio defined as $R = \Sigma_+/\Sigma_-$, and obtain $R=1.68$. In addition, we also calculate the shock strength ($\Psi = M_-/M_+$) which determines the temperature jump across the shock and find $\Psi=1.91$. Notably, when the soft photons from the cooler pre-shock flow interact with the hot electrons of the PSC, they undergo inverse Comptonization, resulting in the production of hard X-rays. This conjecture clearly indicates that the emission of high-energy radiation from black hole sources is inherently influenced by the characteristics of the PSC, $i.e.$, $r_{\rm s}$, $R$ and $\Psi$, respectively. 
	
\begin{figure}
    \begin{center}
        \includegraphics[width=\columnwidth]{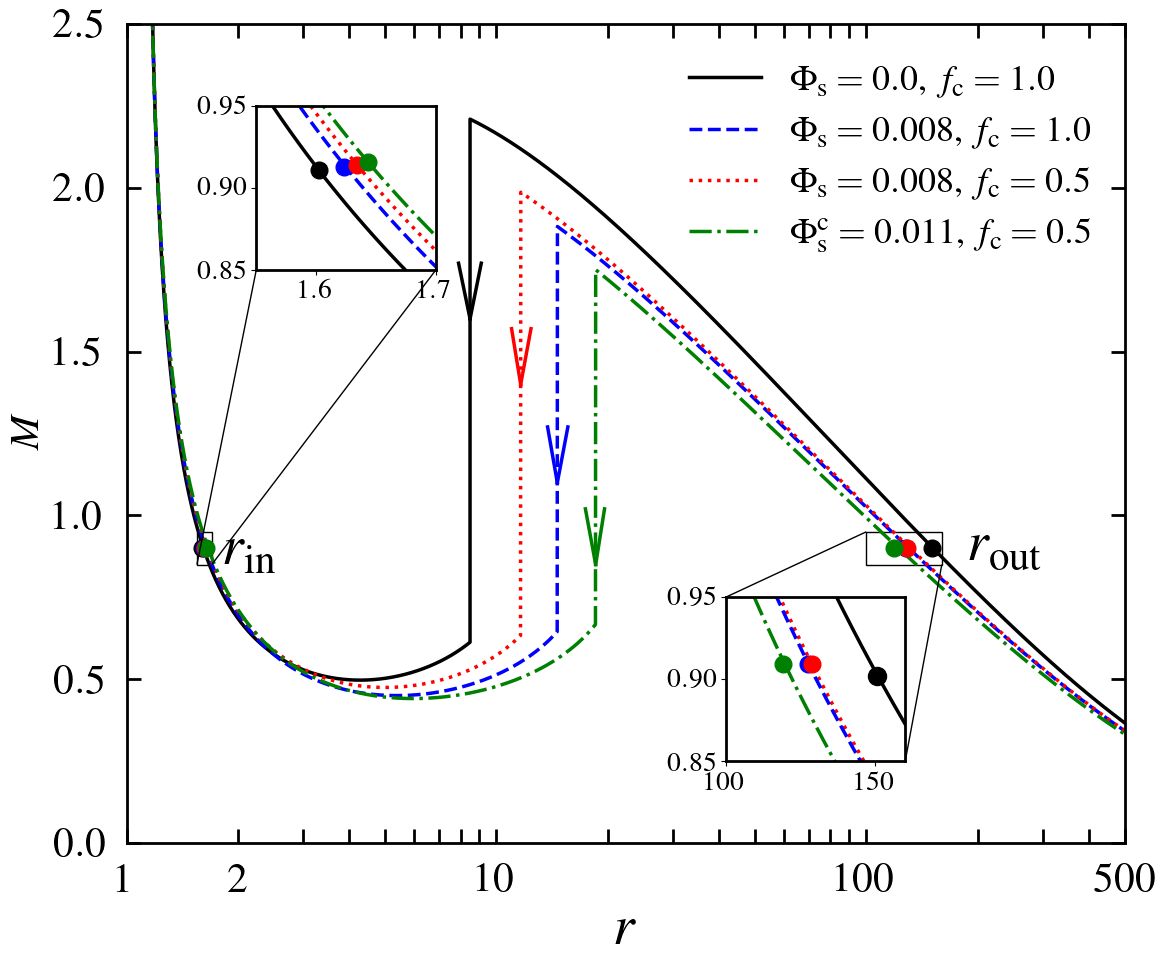}
    \end{center}
    \caption{Plot of Mach number ($M$) with radial coordinate ($r$) for flows injected from $r_{\rm{edge}} = 500$ with $\lambda_{\rm{edge}} = 2.351$, $\mathcal{E_{\rm{edge}}}=1.0015$, and $\alpha=0.01$ around a rapidly rotating black hole of $a_{\rm k}=0.99$. The conduction parameter ($\Phi_{\rm s}$) and cooling parameter ($f_{\rm c}$) are marked in the figure and the corresponding shock radii are obtained as $r_{\rm s}=8.486$ (solid), $11.637$ (dotted), $14.619$ (dashed) and $18.548$ (dot-dashed). Critical points are zoomed at the insets for clarity. See the text for details.
    }
    \label{fig:rs-phi}
\end{figure}

Since the shock radius determines the size of the PSC, it is useful to study the effect of thermal conduction in regulating the shock formation around the rapidly rotating black hole. Towards this, in Fig. \ref{fig:rs-phi}, we illustrate how the shock radius alters due to the increase of the conduction parameter ($\Phi_{\rm s}$) for flows with fixed outer boundary. We inject matter subsonically from $r_{\rm edge}=500$ with $\mathcal{E}_{\rm edge}=1.0015$, $\lambda_{\rm edge}=2.351$, $\alpha =0.01$, and $f_{\rm c}=1.0$ into a black hole of spin $a_{\rm k}=0.99$. For $\Phi_{\rm s}=0.0$, the flow crosses the outer critical point at $r_{\rm out}=150.549$ to become supersonic and experiences a shock transition at $r_{\rm s} = 8.486$ (solid vertical arrow) as the RHCs are satisfied there. Here, we obtain $R=2.65$ and $\Psi=3.60$. As the thermal conduction is turned on ($\Phi_{\rm s} = 0.008$) keeping the other model parameters unchanged, we observe that shock front is pushed outward and it settles down to a larger radius at $r_{\rm s}=14.619$ as shown using dashed vertical arrow. Due to thermal conduction, the local thermal pressure of the accreting flow is increased (see Fig. \ref{fig:MrPhis}), which eventually pushes the shock front outward to maintain the pressure balance across it. For this shocked solution, we obtain $R=2.30$ and $\Psi=2.91$. Next, we turn on parametric cooling by setting $f_{\rm c}=0.5$ and observe that the shock front shifts towards the horizon, with a radius of $r_{\rm s}=11.637$, as shown by the dotted vertical arrow. This is not surprising, as cooling reduces the post-shock pressure, yielding the shock to move inward. When the conduction parameter is further increased to $\Phi_{\rm s} = 0.011$ with cooling ($f_{\rm c} = 0.5$), the shock moves outward to $r_{\rm s} = 18.548$ (as indicated by the dot-dashed vertical arrow), as expected. Beyond a critical value of conduction parameter $\Phi^{\rm cri}_{\rm s} =0.011$, shock disappears as RHCs are not satisfied. These findings clearly indicate that thermal conduction plays a crucial role in shock formation within a dissipative accretion flow around a rotating black hole. We present the model parameters and the obtained shock properties in Table \ref{tab:table-2}.

\begin{figure}
	\begin{center}
		\includegraphics[width=0.9\columnwidth]{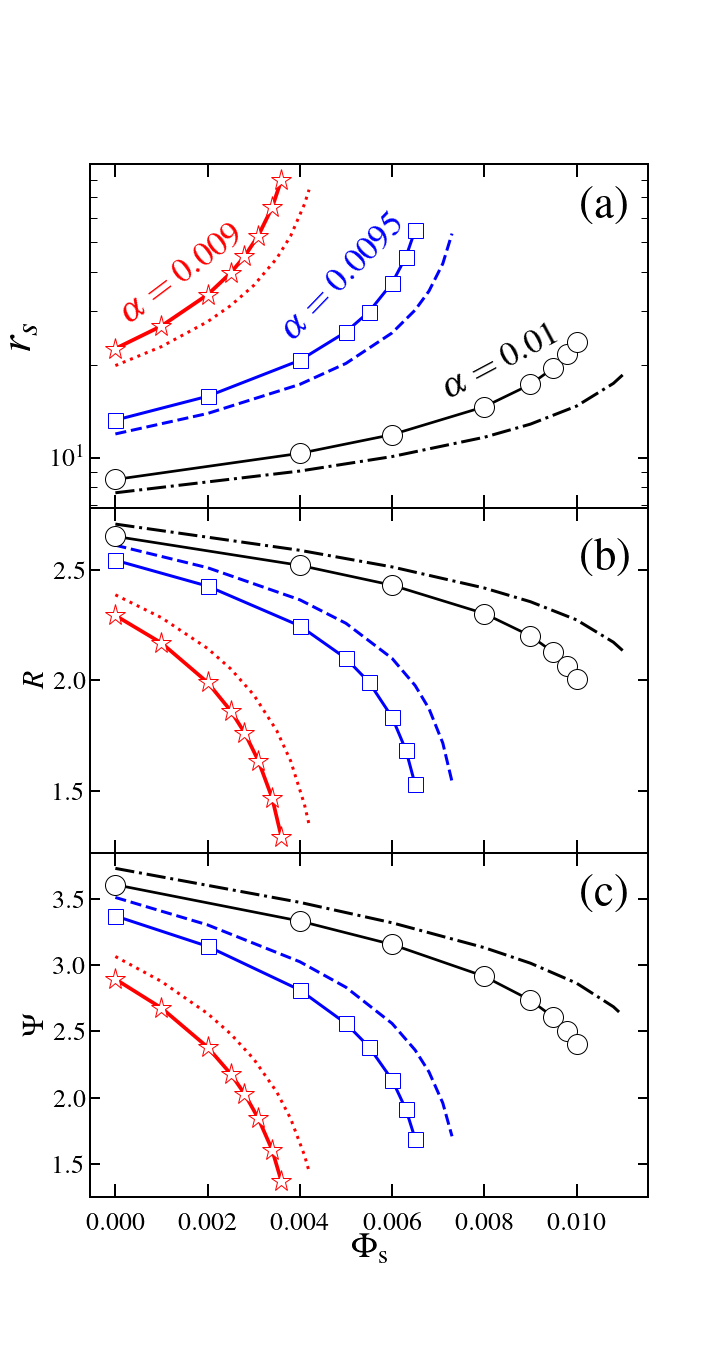}
	\end{center}
	\caption{Variation of (a) shock location $r_{s}$, (b) compression ratio $R$ and (c) shock strength $\Psi$ with conduction parameter $\Phi_{\rm s}$ for different viscosity parameter $\alpha$. Here, flows are injected from the fixed outer edge $r_{\rm edge}=500$ with same $\mathcal{E}_{\rm edge}=1.0015$ and $\lambda_{\rm edge}=2.351$. Asterisks, squares and circles joined with solid lines denote results for $\alpha = 0.009$, $0.0095$ and $0.01$, respectively when cooling is absent ($f_{\rm c}=1.0$). Similarly, in presence of cooling ($f_{\rm c}=0.5$), dotted, dashed and dot-dashed curves represent results for $\alpha = 0.009$, $0.0095$ and $0.01$. See the text for details.
	}
	\label{fig:rsRS-phi}
\end{figure}

\begin{table}
        \centering
        \caption{Details of the shock-induced global accretion solution in presence of thermal conduction. Column $1-9$ refer conduction parameter ($\Phi_{\rm s }$), cooling parameter ($f_{\rm c}$), inner critical point ($r_{\rm in}$), angular momentum at $r_{\rm in}$ ($\lambda_{\rm in}$), outer critical point ($r_{\rm out}$), angular momentum at $r_{\rm out}$ ($\lambda_{\rm out}$), shock radius ($r_{\rm s}$), compression ratio ($R$), and shock strength($\Psi$) for shocked solutions presented in Figure-\ref{fig:rs-phi}. See the text for details.}
        \label{tab:table-2}
        \begin{tabular}{c c c c c c c c c} 
        \hline \hline 
        $\Phi_{\rm{s}}$&$f_{\rm{c}}$&$r_{\textrm{in}}$&$\lambda_{\textrm{in}}$&$r_{\textrm{out}}$&$\lambda_{\textrm{out}}$&$r_{\textrm{s}}$&$R$&$S$ \\
             & & ($r_{\textrm{g}}$) & ($r_{\textrm{g}} c$)& ($r_{\textrm{g}}$)&($r_{\textrm{g}} c$)&($r_{\textrm{g}}$)
            \\ \hline
            0.0& 1.0 &1.602&1.932&150.549&2.037&8.486&2.65&3.60 \\
            0.008&1.0&1.622&1.9106&127.644&2.003&14.619&2.30&2.91 \\
            0.008&0.5&1.633&1.9109&128.759&2.004&11.637&2.41&3.13 \\
            0.011&0.5&1.643&1.901&119.085&1.987&18.548&2.13&2.62 \\
            \hline \\
        \end{tabular}
    \end{table}

In Fig. \ref{fig:rsRS-phi}, we depict the variation of the shock properties, namely shock radius $r_{\rm s}$ (panel a), compression ratio $R$ (panel b), and shock strength $\Psi$ (panel c), as function of conduction parameter $\Phi_{\rm s}$ for different viscosity parameters ($\alpha$). Here, we fix the outer boundary parameters as $\mathcal{E}_{\rm edge}=1.0015$, $\lambda_{\rm edge}=2.351$ at $r_{\rm edge}=500$ in all cases and choose $a_{\rm k}=0.99$. In each panel, the asterisks, squares, and circles connected by solid lines represent results obtained for viscosity parameters $\alpha = 0.009$, $0.0095$, and $0.01$, respectively, with $f_{\rm c} = 1.0$. We observe that standing shocks continue to form across a wide range of conduction parameters ($\Phi_{\rm s}$) in a viscous accretion flow around a rapidly rotating black hole. We also notice that for a fixed $\alpha$, shock front moves further out with the increase of $\Phi_{\rm s}$, as the gas pressure is enhanced due to the increase of thermal conduction. Intriguingly, once $\Phi_{\rm s}$ exceeds a critical value ($\Phi_{\rm s}>\Phi^{\rm cri }_{\rm s}$), the shock disappears because the RHCs are no longer satisfied. Notably, $\Phi^{\rm cri }_{\rm s}$ does not have an universal value as it depends on the other model parameters (see \S 3.4). Moreover, we find that for a given $\Phi_{\rm s}$, when the effect of viscosity is weak, shock forms at larger radii for flows with fixed energy and angular momentum at the outer edge. This happens because lower viscosity effectively diminishes the angular momentum transport outward resulting in a stronger centrifugal barrier and hence shock front settles at larger radii. This findings indicates that shocks are mainly centrifugally supported. Next, we turn on the parametric cooling ($f_c=0.5$) keeping the other model parameters unchanged, and observe that shock forms at relatively smaller radii. The results are shown using dotted, dashed, and dot-dashed curves, which are obtained for $\alpha = 0.009$, $0.0095$ and $0.01$, respectively. Interestingly, $\Phi^{\rm cri}_{\rm s}$ is found to be higher for lower values of $f_c$, irrespective of the viscosity parameter $\alpha$. In panel (b) and (c), we present the variation of the compression ratio $R$ and shock strength $\Psi$ as function of $\Phi_{\rm s}$ for the same set of model parameters used in panel (a). We find that for fixed $\alpha$, both $R$ and $\Psi$ are decreased with the increase of $\Phi_{\rm s}$ regardless of whether cooling is present or absent. We observe a cut-off in $R$ and $\Psi$ when shock ceases to exit for $\Phi_{\rm s}>\Phi^{\rm cri }_{\rm s}$.

\begin{figure}
     \begin{center}
        \includegraphics[width=\columnwidth]{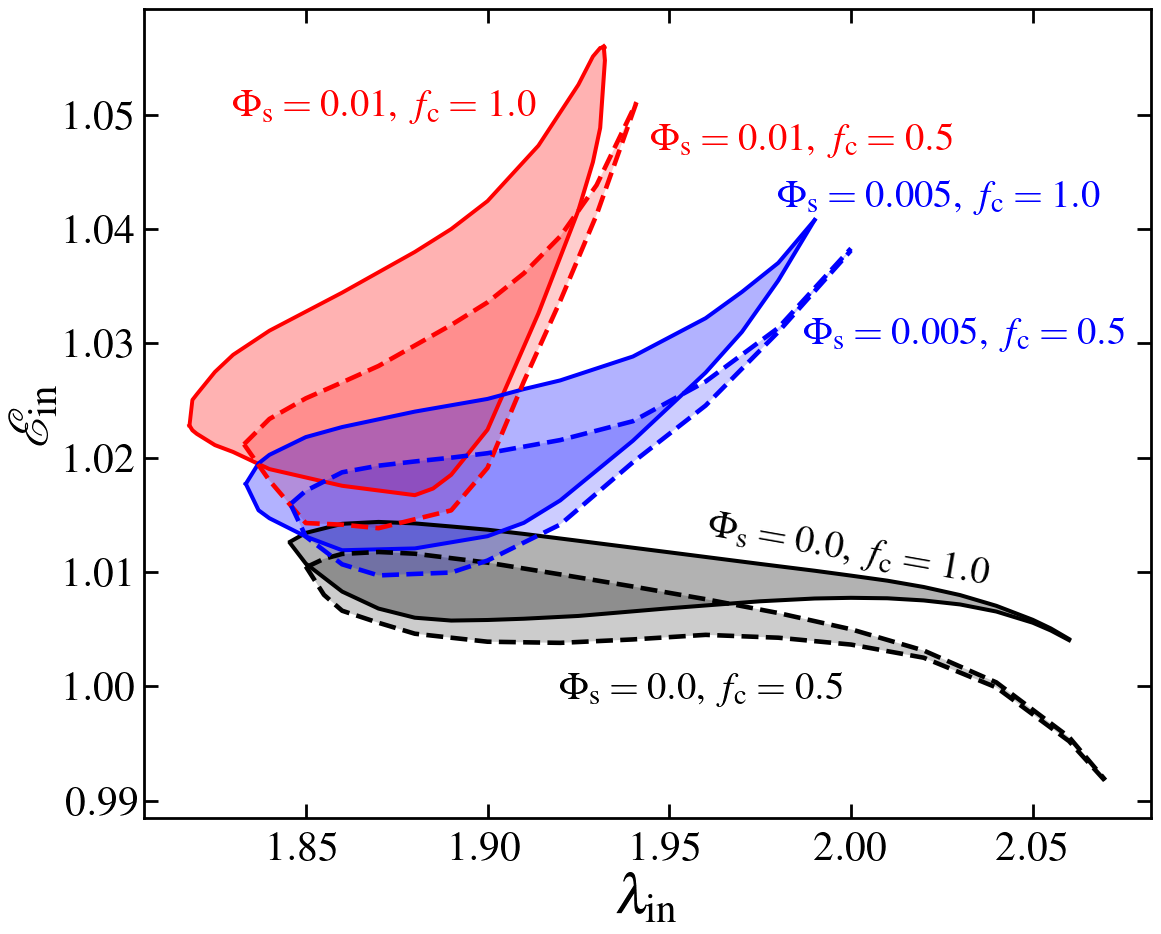}
    \end{center}
    \caption{Modification of the parameter space for standing shock due to different sets of $\Phi_{\rm s}$ and $f_{\rm c}$. Effective regions bounded with solid and dashed curves are obtained in absence and presence of cooling. Here, we choose $\alpha =0.01$ and $a_{\rm k}=0.99$. See the text for details.}
    \label{fig:shock-para}
\end{figure}

\subsection{Shock Parameter Space in energy-angular momentum plane}

As outlined in \S \ref{sec:shock} that shock continues to form in a dissipative accretion flow even in presence of significant thermal conduction. Hence, it is important to investigate the ranges of model parameters that admit shock-induced global accretion solution when thermal conduction is taken into account. To this end, we examine the ranges of energy ($\mathcal{E}_{\rm in}$) and angular momentum ($\lambda_{\rm in}$) of the flow measured at the inner critical point ($r_{\rm in}$) that render shocked solutions. This is done simply because, for a dissipative accretion flow around rapidly rotating black hole ($a_{\rm k}=0.99$), the feasible ranges of the inner critical point and the angular momentum at the inner edge are exceedingly narrow \cite[$r_{\rm in} \lesssim 2$; $1.5 \lesssim \lambda_{\rm in} \lesssim 3$;][]{Chakrabarti-1990,chakrabarti-das-2004}. Accordingly, in Fig. \ref{fig:shock-para}, we separate the region of the parameter space in $\lambda_{\rm in}-\mathcal{E}_{\rm in}$ plane that allows standing shocks in dissipative accretion flow and study how the shock parameter space alters for increasing conduction and cooling parameters ($\Phi_{\rm s}$, $f_{\rm c}$). Here, we choose $\alpha = 0.01$ and $a_{\rm k}=0.99$. In the figure, conduction parameter ($\Phi_{\rm s}$) and cooling parameter ($f_{\rm c}$) are marked, where the regions bounded with solid and dashed curves are obtained in absence ($f_{\rm c}=1.0$) and presence ($f_{\rm c}=0.5$) of cooling. We observe that increasing the conduction parameter ($\Phi_{\rm s}$) in the absence of parametric cooling shifts the effective region of the parameter space toward lower angular momentum and higher energy domains. This shift is expected because higher thermal conduction raises the gas pressure, which in turn enhances angular momentum transport (see equation (\ref{eqn:vis-strees})). When cooling is activated ($f_{\rm c}=0.5$) within the disk, the energy of the flow decreases, causing the shock parameter space to shift toward lower energy regions, as illustrated by the dashed curve.

\subsection{Critical Conduction Parameter for Standing Shock}
\label{sec:Cri-Phis}

\begin{figure*}
    \begin{center}
    \includegraphics[width=\columnwidth]{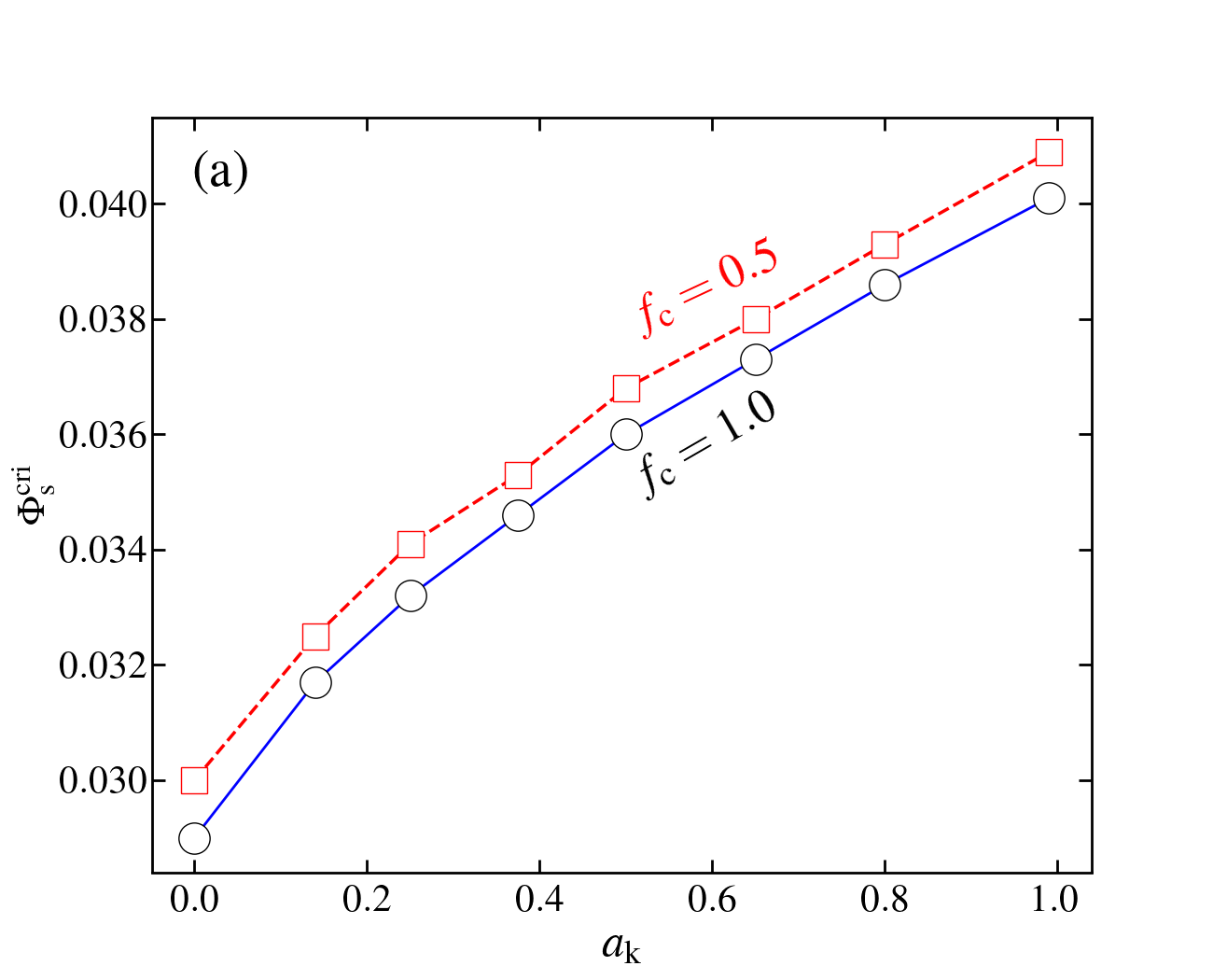}
    \includegraphics[width=\columnwidth]{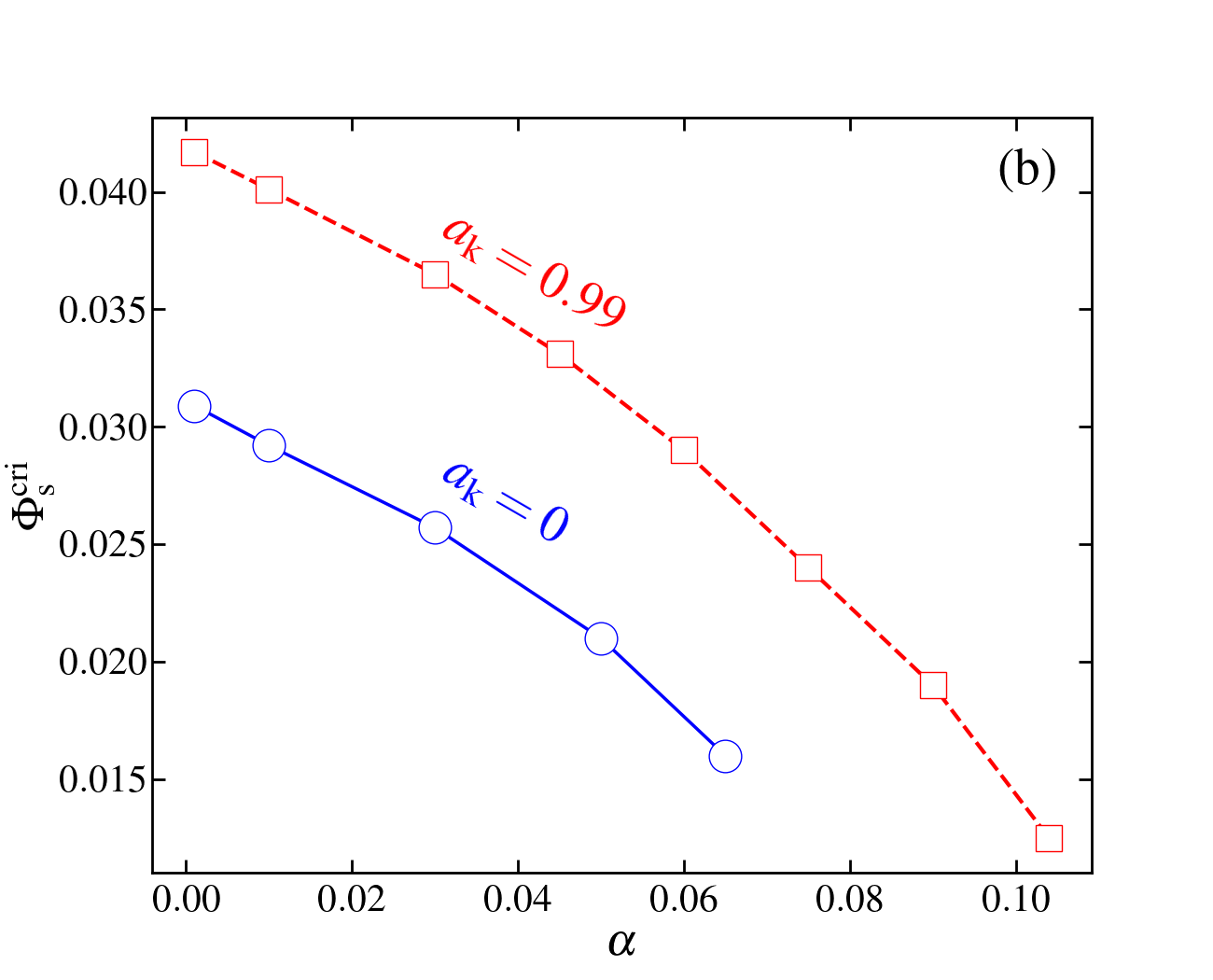}
    \end{center}
    \caption{Panel(a): Variation of $\Phi_{\rm s}^{\rm cri}$ with black hole spin ($a_{\rm k}$) for flows with $\alpha =0.01$. Open circles and squares joined with solid and dashed lines denote results in absence ($f_{\rm c}=1.0$) and presence ($f_{\rm c}=0.5$) of cooling. Panel (b): Variation of $\Phi_{\rm s}^{\rm cri}$ with viscosity parameter ($\alpha$) for different $a_{\rm k}$ values. Open circles and squares joined with solid and dashed lines are for weakly rotating ($a_{\rm k}=0.0$) and rapidly rotating ($a_{\rm k}=0.99$) black holes. Here, we choose $f_{\rm c}=1.0$. See the text for details.
    }
    \label{fig:Cri-phis}
\end{figure*}

Meanwhile, we have pointed out that when the conduction parameter ($\Phi_{\rm s}$) exceeds a critical value ($\Phi_{\rm s} > \Phi^{\rm cri}_{\rm s}$), the accretion flow doesn't sustain shock waves because RHCs become unfavorable. Indeed, $\Phi^{\rm cri}_{\rm s}$ is largely depends on the other model parameters. Therefore, we determine $\Phi^{\rm cri}_{\rm s}$ as function of black hole spin ($a_{\rm k}$), considering both cases with and without cooling. Here, we keep the viscosity parameter fixed as $\alpha = 0.01$, and freely vary energy ($\mathcal{E}_{\rm in}$) and angular momentum ($\lambda_{\rm in}$) at the inner critical point ($r_{\rm in}$). The obtained results are presented in Fig. \ref{fig:Cri-phis}a, where open circles joined with solid lines denote the variation of $\Phi^{\rm cri}_{\rm s}$ corresponding to $f_{\rm c}=1.0$, whereas open squares connected with dotted lines refers the same for $f_{\rm c}=0.5$. Figure evidently suggests that shock-induced global accretion solutions continue to exist for wide range of $\Phi_{\rm s}$, and $\Phi^{\rm cri}_{\rm s}$ increases with $a_{\rm k}$. Moreover, our findings reveals that the critical conduction parameter $\Phi^{\rm cri}_{\rm s}$ consistently exhibits higher values when cooling is active within the flow. In particular, we find that in absence of cooling $f_{\rm c}=1.0$, $\Phi^{\rm cri}_{\rm s}=0.029$ ($0.0401$) for non-rotating black hole of $a_{\rm k}=0.0$ (rapidly rotating black hole of $a_{\rm k}=0.99$). Similarly, when cooling is present $f_{\rm c}=0.5$, we obtain $\Phi^{\rm cri}_{\rm s}=0.03$ ($0.0409$) for $a_{\rm k}=0.0$ ($0.99$). This findings suggest that dissipative accretion flows around rapidly rotating black holes continue to harbour shocks even in presence of higher thermal conduction and vice versa. 

Furthermore, we investigate how thermal conduction influences the possibility of shock transitions in dissipative accretion flows characterized by different viscosity. In doing so, we follow the same approach delineated above, and obtain $\Phi^{\rm cri}_{\rm s}$ for increasing values of $\alpha$ keeping the black hole spin ($a_{\rm k}$) fixed. Here, we choose $f_{\rm c}=1.0$ disregarding any cooling effects. The obtained results are depicted in Fig. \ref{fig:Cri-phis}b, where open circles connected with solid line are for $a_{\rm k}=0.0$, and open squares joined with dashed lines are for $a_{\rm k}=0.99$. We observe that $\Phi^{\rm cri}_{\rm s}$ monotonically decreases with the increase of $\alpha$ irrespective to $a_{\rm k}$ values. This happens due to the combined effects of thermal conduction and viscosity that maintains the level of dissipation favourable to trigger the shock transition inside the accretion flow around rotating black holes. We also observe that shocks disappear at relatively lower viscosity ($\alpha=0.065$) for weakly rotating black holes ($a_{\rm k} \rightarrow 0$), while for rapidly rotating ($a_{\rm k} \rightarrow 1$) black holes, shocks persist even at higher values as $\alpha=0.104$.

\section{CONCLUSION}

In this paper, we address for the first time the effects of thermal conduction on the shock-induced viscous advective accretion flows around weakly rotating ($a_{\rm k} \rightarrow 0$) as well as rapidly rotating ($a_{\rm k} \rightarrow 1$) black holes in presence of cooling. By adopting a pseudo-Kerr effective potential that satisfactorily describes the spacetime geometry around rotating black hole, and incorporating the relativistic equation of state (REoS), we solve the governing flow equations and find that thermal conduction plays an important role in triggering the shock transition and influencing its characteristics, namely shock location ($r_{\rm s}$), compression ratio ($R$) and shock strength ($\Psi$), respectively. Below, we summarize the key findings of this work point wise.

\begin{itemize}

    \item We self-consistently calculate the global transonic accretion solutions around rotating black holes taking into account the effects of thermal conduction. We observe that, depending on the conduction parameter ($\Phi_{\rm s}$), the accretion flow with a fixed outer boundary either passes through the inner ($r_{\rm in}$) or the outer ($r_{\rm out}$) critical point before entering into the black hole (see Fig. \ref{fig:MrPhis}).

    \item We find that accretion solutions passing through $r_{\rm out}$ undergo a shock transition, provided the Rankine-Hugoniot conditions (RHCs) are satisfied \cite[]{Landau-Lifshitz1959}. This happens because shocked solutions possess higher entropy compared to shock-free solutions \cite[]{Becker-Kazanas2001}(see Fig. \ref{fig:shock}). We also observe that for a given set of model parameters ($\mathcal{E}$, $\lambda$, $\alpha$, and $a_{\rm k}$), an increase in the conduction parameter causes the shock to form farther from the black hole. When cooling is turned on, the shock front moves inward as the post-shock flow is cooled down more efficiently, leading to a reduction in thermal pressure and causing the shock front to settle down at smaller radii (see Fig. \ref{fig:rs-phi}).

    \item The properties of the post-shock flow ($i.e.$, PSC) - such as its size, temperature, and density - explicitly depend  on the shock location ($r_{\rm s}$), compression ratio ($R$) and shock strength ($\Psi$). We observe that these quantities are eventually regulated by thermal conduction (see Fig. \ref{fig:rsRS-phi}). Indeed, due to shock transition, post-shock flow (PSC) becomes hot and puffed up containing swarm of hot electrons. When soft photons from the pre-shock flow are intercepted at the PSC, they are reprocessed via inverse Comptonization processes to produce hard X-ray radiations. These high energy radiations are commonly observed from black hole X-ray binaries \cite[]{chakrabarti-titarchuk-1995,nandi-etal-2012,Iyer-etal2015,Nandi-etal2018,Majumder-etal2023}. Since PSC properties are regulated by $\Phi_{\rm s}$, we argue that thermal conduction seems to play important role in influencing the spectral properties of black holes.
    
    \item Furthermore, we find that shock-induced accretion solutions are not isolated solutions, rather they are consistently found over a wide range of model parameters ($\mathcal{E}_{\rm in}$, $\lambda_{\rm in}$, $\alpha$, $\Phi_{\rm s}$ and $a_{\rm k}$). Accordingly, for different $\Phi_{\rm s}$ and $f_{\rm c}$, we identify the effective region of the parameter space spanned by $\lambda_{\rm in}$ and $\mathcal{E}_{\rm in}$ that allows transonic global accretion solutions containing shocks around rapidly rotating black holes (see Fig. \ref{fig:shock-para}). We observe that parameter space alters due to the increase of conduction parameter ($\Phi_{\rm s}$) and cooling parameter ($f_{\rm c}$). It is important to note that thermal conduction and cooling have opposing effects on determining the parameter space for stationary shock waves.

    \item We calculate the critical conduction parameter ($\Phi_{\rm s}^{\rm cri}$), beyond which shocked accretion solutions no longer exist. We find that $\Phi_{\rm s}^{\rm cri}$ is correlated with $a_{\rm k}$, where, in absence of cooling, $\Phi_{\rm s}^{\rm cri} = 0.0401$ for a rapidly rotating black hole ($a_{\rm k} \rightarrow 1$) and $\Phi_{\rm s}^{\rm cri} = 0.029$ for a non-rotating black hole ($a_{\rm k} = 0$). Additionally, we observe that $\Phi_{\rm s}^{\rm cri}$ is higher when cooling is present compared to the scenario without cooling. Furthermore, for a fixed $a_{\rm k}$, $\Phi_{\rm s}^{\rm cri}$ is seen to decrease monotonically with the increase of viscosity parameter $\alpha$ (see Fig. \ref{fig:Cri-phis}).
        
\end{itemize}

Finally, we mention the limitations of this study, which is developed based on certain assumptions. For simplicity, we use parametric cooling \cite[]{Narayan-Yi1994} instead of incorporating realistic radiative processes, such as bremsstrahlung, synchrotron, and Compton cooling. We neglect the effects of magnetic fields and do not account for mass loss from the accretion disk. Indeed, all these physical processes are relevant in the context of the accretion disc dynamics. However. we argue that the key findings of this study are expected to remain qualitatively consistent despite these approximations.

\section*{Data Availability}

The data underlying this paper will be available with reasonable request.

\section*{Acknowledgements}

Authors thank the Department of Physics, IIT Guwahati, India for providing the infrastructural support to carry out this work.


\begin{thebibliography}{78}
\expandafter\ifx\csname natexlab\endcsname\relax\def\natexlab#1{#1}\fi
\expandafter\ifx\csname bibnamefont\endcsname\relax
  \def\bibnamefont#1{#1}\fi
\expandafter\ifx\csname bibfnamefont\endcsname\relax
  \def\bibfnamefont#1{#1}\fi
\expandafter\ifx\csname citenamefont\endcsname\relax
  \def\citenamefont#1{#1}\fi
\expandafter\ifx\csname url\endcsname\relax
  \def\url#1{\texttt{#1}}\fi
\expandafter\ifx\csname urlprefix\endcsname\relax\def\urlprefix{URL }\fi
\providecommand{\bibinfo}[2]{#2}
\providecommand{\eprint}[2][]{\url{#2}}

\bibitem[{\citenamefont{{Frank} et~al.}(2002)\citenamefont{{Frank}, {King}, and
  {Raine}}}]{Frank-etal2002}
\bibinfo{author}{\bibfnamefont{J.}~\bibnamefont{{Frank}}},
  \bibinfo{author}{\bibfnamefont{A.}~\bibnamefont{{King}}}, \bibnamefont{and}
  \bibinfo{author}{\bibfnamefont{D.~J.} \bibnamefont{{Raine}}},
  \emph{\bibinfo{title}{{Accretion Power in Astrophysics: Third Edition}}}
  (\bibinfo{publisher}{Cambridge, UK: Cambridge University Press},
  \bibinfo{year}{2002}).

\bibitem[{\citenamefont{{Shakura} and {Sunyaev}}(1973)}]{shakura-sunyaev-1973}
\bibinfo{author}{\bibfnamefont{N.~I.} \bibnamefont{{Shakura}}}
  \bibnamefont{and} \bibinfo{author}{\bibfnamefont{R.~A.}
  \bibnamefont{{Sunyaev}}}, \bibinfo{journal}{Astronomy and Astrophysics}
  \textbf{\bibinfo{volume}{500}}, \bibinfo{pages}{33} (\bibinfo{year}{1973}).

\bibitem[{\citenamefont{{Novikov} and {Thorne}}(1973)}]{Novikov-Thorne1973}
\bibinfo{author}{\bibfnamefont{I.~D.} \bibnamefont{{Novikov}}}
  \bibnamefont{and} \bibinfo{author}{\bibfnamefont{K.~S.}
  \bibnamefont{{Thorne}}}, in \emph{\bibinfo{booktitle}{Black Holes (Les Astres
  Occlus)}} (\bibinfo{publisher}{New York: Gordon \& Breach},
  \bibinfo{year}{1973}), pp. \bibinfo{pages}{343--450}.

\bibitem[{\citenamefont{{Abramowicz} et~al.}(1988)\citenamefont{{Abramowicz},
  {Czerny}, {Lasota}, and {Szuszkiewicz}}}]{Abramowicz-etal1988}
\bibinfo{author}{\bibfnamefont{M.~A.} \bibnamefont{{Abramowicz}}},
  \bibinfo{author}{\bibfnamefont{B.}~\bibnamefont{{Czerny}}},
  \bibinfo{author}{\bibfnamefont{J.~P.} \bibnamefont{{Lasota}}},
  \bibnamefont{and}
  \bibinfo{author}{\bibfnamefont{E.}~\bibnamefont{{Szuszkiewicz}}},
  \bibinfo{journal}{\apj} \textbf{\bibinfo{volume}{332}}, \bibinfo{pages}{646}
  (\bibinfo{year}{1988}).

\bibitem[{\citenamefont{{Narayan} and {Yi}}(1994)}]{Narayan-Yi1994}
\bibinfo{author}{\bibfnamefont{R.}~\bibnamefont{{Narayan}}} \bibnamefont{and}
  \bibinfo{author}{\bibfnamefont{I.}~\bibnamefont{{Yi}}},
  \bibinfo{journal}{\apjl} \textbf{\bibinfo{volume}{428}}, \bibinfo{pages}{L13}
  (\bibinfo{year}{1994}), \eprint{astro-ph/9403052}.

\bibitem[{\citenamefont{{Narayan} and {Yi}}(1995)}]{Narayan-Yi1995}
\bibinfo{author}{\bibfnamefont{R.}~\bibnamefont{{Narayan}}} \bibnamefont{and}
  \bibinfo{author}{\bibfnamefont{I.}~\bibnamefont{{Yi}}},
  \bibinfo{journal}{\apj} \textbf{\bibinfo{volume}{452}}, \bibinfo{pages}{710}
  (\bibinfo{year}{1995}), \eprint{astro-ph/9411059}.

\bibitem[{\citenamefont{{Esin} et~al.}(1997)\citenamefont{{Esin}, {McClintock},
  and {Narayan}}}]{Esin-etal1997}
\bibinfo{author}{\bibfnamefont{A.~A.} \bibnamefont{{Esin}}},
  \bibinfo{author}{\bibfnamefont{J.~E.} \bibnamefont{{McClintock}}},
  \bibnamefont{and}
  \bibinfo{author}{\bibfnamefont{R.}~\bibnamefont{{Narayan}}},
  \bibinfo{journal}{\apj} \textbf{\bibinfo{volume}{489}}, \bibinfo{pages}{865}
  (\bibinfo{year}{1997}), \eprint{astro-ph/9705237}.

\bibitem[{\citenamefont{{Hameury} et~al.}(1997)\citenamefont{{Hameury},
  {Lasota}, {McClintock}, and {Narayan}}}]{Hameury-etal1997}
\bibinfo{author}{\bibfnamefont{J.~M.} \bibnamefont{{Hameury}}},
  \bibinfo{author}{\bibfnamefont{J.~P.} \bibnamefont{{Lasota}}},
  \bibinfo{author}{\bibfnamefont{J.~E.} \bibnamefont{{McClintock}}},
  \bibnamefont{and}
  \bibinfo{author}{\bibfnamefont{R.}~\bibnamefont{{Narayan}}},
  \bibinfo{journal}{\apj} \textbf{\bibinfo{volume}{489}}, \bibinfo{pages}{234}
  (\bibinfo{year}{1997}), \eprint{astro-ph/9703095}.

\bibitem[{\citenamefont{{Yuan} and {Cui}}(2005)}]{Yuan-Cui2005}
\bibinfo{author}{\bibfnamefont{F.}~\bibnamefont{{Yuan}}} \bibnamefont{and}
  \bibinfo{author}{\bibfnamefont{W.}~\bibnamefont{{Cui}}},
  \bibinfo{journal}{\apj} \textbf{\bibinfo{volume}{629}}, \bibinfo{pages}{408}
  (\bibinfo{year}{2005}), \eprint{astro-ph/0411770}.

\bibitem[{\citenamefont{{Liu} et~al.}(2011)\citenamefont{{Liu}, {Done}, and
  {Taam}}}]{Liu-etal2011}
\bibinfo{author}{\bibfnamefont{B.~F.} \bibnamefont{{Liu}}},
  \bibinfo{author}{\bibfnamefont{C.}~\bibnamefont{{Done}}}, \bibnamefont{and}
  \bibinfo{author}{\bibfnamefont{R.~E.} \bibnamefont{{Taam}}},
  \bibinfo{journal}{\apj} \textbf{\bibinfo{volume}{726}}, \bibinfo{eid}{10}
  (\bibinfo{year}{2011}), \eprint{1011.2580}.

\bibitem[{\citenamefont{{Reynolds} et~al.}(1996)\citenamefont{{Reynolds}, {Di
  Matteo}, {Fabian}, {Hwang}, and {Canizares}}}]{Reynolds-etal1996}
\bibinfo{author}{\bibfnamefont{C.~S.} \bibnamefont{{Reynolds}}},
  \bibinfo{author}{\bibfnamefont{T.}~\bibnamefont{{Di Matteo}}},
  \bibinfo{author}{\bibfnamefont{A.~C.} \bibnamefont{{Fabian}}},
  \bibinfo{author}{\bibfnamefont{U.}~\bibnamefont{{Hwang}}}, \bibnamefont{and}
  \bibinfo{author}{\bibfnamefont{C.~R.} \bibnamefont{{Canizares}}},
  \bibinfo{journal}{\mnras} \textbf{\bibinfo{volume}{283}},
  \bibinfo{pages}{L111} (\bibinfo{year}{1996}), \eprint{astro-ph/9610097}.

\bibitem[{\citenamefont{{Manmoto} et~al.}(1997)\citenamefont{{Manmoto},
  {Mineshige}, and {Kusunose}}}]{Manmoto-etal1997}
\bibinfo{author}{\bibfnamefont{T.}~\bibnamefont{{Manmoto}}},
  \bibinfo{author}{\bibfnamefont{S.}~\bibnamefont{{Mineshige}}},
  \bibnamefont{and}
  \bibinfo{author}{\bibfnamefont{M.}~\bibnamefont{{Kusunose}}},
  \bibinfo{journal}{\apj} \textbf{\bibinfo{volume}{489}}, \bibinfo{pages}{791}
  (\bibinfo{year}{1997}), \eprint{astro-ph/9708234}.

\bibitem[{\citenamefont{{Yuan} and {Narayan}}(2014)}]{Yuan-Narayan2014}
\bibinfo{author}{\bibfnamefont{F.}~\bibnamefont{{Yuan}}} \bibnamefont{and}
  \bibinfo{author}{\bibfnamefont{R.}~\bibnamefont{{Narayan}}},
  \bibinfo{journal}{\araa} \textbf{\bibinfo{volume}{52}}, \bibinfo{pages}{529}
  (\bibinfo{year}{2014}), \eprint{1401.0586}.

\bibitem[{\citenamefont{{Younes} et~al.}(2019)\citenamefont{{Younes}, {Ptak},
  {Ho}, {Xie}, {Terasima}, {Yuan}, {Huppenkothen}, and
  {Yukita}}}]{Younes-etal2019}
\bibinfo{author}{\bibfnamefont{G.}~\bibnamefont{{Younes}}},
  \bibinfo{author}{\bibfnamefont{A.}~\bibnamefont{{Ptak}}},
  \bibinfo{author}{\bibfnamefont{L.~C.} \bibnamefont{{Ho}}},
  \bibinfo{author}{\bibfnamefont{F.-G.} \bibnamefont{{Xie}}},
  \bibinfo{author}{\bibfnamefont{Y.}~\bibnamefont{{Terasima}}},
  \bibinfo{author}{\bibfnamefont{F.}~\bibnamefont{{Yuan}}},
  \bibinfo{author}{\bibfnamefont{D.}~\bibnamefont{{Huppenkothen}}},
  \bibnamefont{and} \bibinfo{author}{\bibfnamefont{M.}~\bibnamefont{{Yukita}}},
  \bibinfo{journal}{\apj} \textbf{\bibinfo{volume}{870}}, \bibinfo{eid}{73}
  (\bibinfo{year}{2019}), \eprint{1811.10657}.

\bibitem[{\citenamefont{{Ichimaru}}(1977)}]{Ichimaru1977}
\bibinfo{author}{\bibfnamefont{S.}~\bibnamefont{{Ichimaru}}},
  \bibinfo{journal}{\apj} \textbf{\bibinfo{volume}{214}}, \bibinfo{pages}{840}
  (\bibinfo{year}{1977}).

\bibitem[{\citenamefont{{Tanaka} and {Menou}}(2006)}]{tanaka-menou-2006}
\bibinfo{author}{\bibfnamefont{T.}~\bibnamefont{{Tanaka}}} \bibnamefont{and}
  \bibinfo{author}{\bibfnamefont{K.}~\bibnamefont{{Menou}}},
  \bibinfo{journal}{\apj} \textbf{\bibinfo{volume}{649}}, \bibinfo{pages}{345}
  (\bibinfo{year}{2006}), \eprint{astro-ph/0604509}.

\bibitem[{\citenamefont{{Johnson} and {Quataert}}(2007)}]{Johnson-Quataert2007}
\bibinfo{author}{\bibfnamefont{B.~M.} \bibnamefont{{Johnson}}}
  \bibnamefont{and}
  \bibinfo{author}{\bibfnamefont{E.}~\bibnamefont{{Quataert}}},
  \bibinfo{journal}{\apj} \textbf{\bibinfo{volume}{660}}, \bibinfo{pages}{1273}
  (\bibinfo{year}{2007}), \eprint{astro-ph/0608467}.

\bibitem[{\citenamefont{{Shadmehri}}(2008)}]{shadmehri-2008}
\bibinfo{author}{\bibfnamefont{M.}~\bibnamefont{{Shadmehri}}},
  \bibinfo{journal}{Astrophysics and Space Science}
  \textbf{\bibinfo{volume}{317}}, \bibinfo{pages}{201} (\bibinfo{year}{2008}),
  \eprint{0808.0245}.

\bibitem[{\citenamefont{{Faghei}}(2012)}]{Faghei2012}
\bibinfo{author}{\bibfnamefont{K.}~\bibnamefont{{Faghei}}},
  \bibinfo{journal}{\mnras} \textbf{\bibinfo{volume}{420}},
  \bibinfo{pages}{118} (\bibinfo{year}{2012}), \eprint{1111.3569}.

\bibitem[{\citenamefont{{Khajenabi} and
  {Shadmehri}}(2013)}]{Khajenabi-Shadmehri-2013}
\bibinfo{author}{\bibfnamefont{F.}~\bibnamefont{{Khajenabi}}} \bibnamefont{and}
  \bibinfo{author}{\bibfnamefont{M.}~\bibnamefont{{Shadmehri}}},
  \bibinfo{journal}{\mnras} \textbf{\bibinfo{volume}{436}},
  \bibinfo{pages}{2666} (\bibinfo{year}{2013}), \eprint{1309.5710}.

\bibitem[{\citenamefont{{Ghoreyshi} and
  {Shadmehri}}(2020)}]{ghoreyshi-shadmehri-2020}
\bibinfo{author}{\bibfnamefont{S.~M.} \bibnamefont{{Ghoreyshi}}}
  \bibnamefont{and}
  \bibinfo{author}{\bibfnamefont{M.}~\bibnamefont{{Shadmehri}}},
  \bibinfo{journal}{Monthly Notices of the Royal Astronomical Society}
  \textbf{\bibinfo{volume}{493}}, \bibinfo{pages}{5107} (\bibinfo{year}{2020}),
  \eprint{2003.04752}.

\bibitem[{\citenamefont{{Mosallanezhad}
  et~al.}(2021)\citenamefont{{Mosallanezhad}, {Zeraatgari}, {Mei}, and
  {Bu}}}]{Mosallanezhad-etal2021}
\bibinfo{author}{\bibfnamefont{A.}~\bibnamefont{{Mosallanezhad}}},
  \bibinfo{author}{\bibfnamefont{F.~Z.} \bibnamefont{{Zeraatgari}}},
  \bibinfo{author}{\bibfnamefont{L.}~\bibnamefont{{Mei}}}, \bibnamefont{and}
  \bibinfo{author}{\bibfnamefont{D.-F.} \bibnamefont{{Bu}}},
  \bibinfo{journal}{\apj} \textbf{\bibinfo{volume}{909}}, \bibinfo{eid}{140}
  (\bibinfo{year}{2021}), \eprint{2101.08006}.

\bibitem[{\citenamefont{{Bu} et~al.}(2016)\citenamefont{{Bu}, {Wu}, and
  {Yuan}}}]{Bu-etal2016}
\bibinfo{author}{\bibfnamefont{D.-F.} \bibnamefont{{Bu}}},
  \bibinfo{author}{\bibfnamefont{M.-C.} \bibnamefont{{Wu}}}, \bibnamefont{and}
  \bibinfo{author}{\bibfnamefont{Y.-F.} \bibnamefont{{Yuan}}},
  \bibinfo{journal}{\mnras} \textbf{\bibinfo{volume}{459}},
  \bibinfo{pages}{746} (\bibinfo{year}{2016}), \eprint{1603.07407}.

\bibitem[{\citenamefont{{Mitra} et~al.}(2023)\citenamefont{{Mitra},
  {Ghoreyshi}, {Mosallanezhad}, {Abbassi}, and {Das}}}]{mitra-etal-2023}
\bibinfo{author}{\bibfnamefont{S.}~\bibnamefont{{Mitra}}},
  \bibinfo{author}{\bibfnamefont{S.~M.} \bibnamefont{{Ghoreyshi}}},
  \bibinfo{author}{\bibfnamefont{A.}~\bibnamefont{{Mosallanezhad}}},
  \bibinfo{author}{\bibfnamefont{S.}~\bibnamefont{{Abbassi}}},
  \bibnamefont{and} \bibinfo{author}{\bibfnamefont{S.}~\bibnamefont{{Das}}},
  \bibinfo{journal}{Monthly Notices of the Royal Astronomical Society}
  \textbf{\bibinfo{volume}{523}}, \bibinfo{pages}{4431} (\bibinfo{year}{2023}),
  \eprint{2306.02453}.

\bibitem[{\citenamefont{{Mitra} and {Das}}(2024)}]{Mitra-Das2024}
\bibinfo{author}{\bibfnamefont{S.}~\bibnamefont{{Mitra}}} \bibnamefont{and}
  \bibinfo{author}{\bibfnamefont{S.}~\bibnamefont{{Das}}},
  \bibinfo{journal}{arXiv e-prints} \bibinfo{eid}{arXiv:2405.16326}
  (\bibinfo{year}{2024}), \eprint{2405.16326}.

\bibitem[{\citenamefont{{Fukue}}(1987)}]{Fukue1987}
\bibinfo{author}{\bibfnamefont{J.}~\bibnamefont{{Fukue}}},
  \bibinfo{journal}{\pasj} \textbf{\bibinfo{volume}{39}}, \bibinfo{pages}{309}
  (\bibinfo{year}{1987}).

\bibitem[{\citenamefont{{Chakrabarti}}(1989)}]{Chakrabarti1989}
\bibinfo{author}{\bibfnamefont{S.~K.} \bibnamefont{{Chakrabarti}}},
  \bibinfo{journal}{\apj} \textbf{\bibinfo{volume}{347}}, \bibinfo{pages}{365}
  (\bibinfo{year}{1989}).

\bibitem[{\citenamefont{{Chakrabarti}}(1996)}]{chakrabarti-1996}
\bibinfo{author}{\bibfnamefont{S.~K.} \bibnamefont{{Chakrabarti}}},
  \bibinfo{journal}{\apj} \textbf{\bibinfo{volume}{464}}, \bibinfo{pages}{664}
  (\bibinfo{year}{1996}), \eprint{astro-ph/9606145}.

\bibitem[{\citenamefont{{Das} et~al.}(2001)\citenamefont{{Das},
  {Chattopadhyay}, and {Chakrabarti}}}]{Das-etal2001}
\bibinfo{author}{\bibfnamefont{S.}~\bibnamefont{{Das}}},
  \bibinfo{author}{\bibfnamefont{I.}~\bibnamefont{{Chattopadhyay}}},
  \bibnamefont{and} \bibinfo{author}{\bibfnamefont{S.~K.}
  \bibnamefont{{Chakrabarti}}}, \bibinfo{journal}{\apj}
  \textbf{\bibinfo{volume}{557}}, \bibinfo{pages}{983} (\bibinfo{year}{2001}),
  \eprint{astro-ph/0107046}.

\bibitem[{\citenamefont{{Das}}(2007)}]{das-2007}
\bibinfo{author}{\bibfnamefont{S.}~\bibnamefont{{Das}}},
  \bibinfo{journal}{\mnras} \textbf{\bibinfo{volume}{376}},
  \bibinfo{pages}{1659} (\bibinfo{year}{2007}), \eprint{astro-ph/0610651}.

\bibitem[{\citenamefont{{Becker} and {Kazanas}}(2001)}]{Becker-Kazanas2001}
\bibinfo{author}{\bibfnamefont{P.~A.} \bibnamefont{{Becker}}} \bibnamefont{and}
  \bibinfo{author}{\bibfnamefont{D.}~\bibnamefont{{Kazanas}}},
  \bibinfo{journal}{The Astrophysical Journal} \textbf{\bibinfo{volume}{546}},
  \bibinfo{pages}{429} (\bibinfo{year}{2001}), \eprint{astro-ph/0101020}.

\bibitem[{\citenamefont{{Yang} and {Kafatos}}(1995)}]{Yang-Kafatos1995}
\bibinfo{author}{\bibfnamefont{R.}~\bibnamefont{{Yang}}} \bibnamefont{and}
  \bibinfo{author}{\bibfnamefont{M.}~\bibnamefont{{Kafatos}}},
  \bibinfo{journal}{\aap} \textbf{\bibinfo{volume}{295}}, \bibinfo{pages}{238}
  (\bibinfo{year}{1995}).

\bibitem[{\citenamefont{{Lu} et~al.}(1999)\citenamefont{{Lu}, {Gu}, and
  {Yuan}}}]{lu-etal-1999}
\bibinfo{author}{\bibfnamefont{J.-F.} \bibnamefont{{Lu}}},
  \bibinfo{author}{\bibfnamefont{W.-M.} \bibnamefont{{Gu}}}, \bibnamefont{and}
  \bibinfo{author}{\bibfnamefont{F.}~\bibnamefont{{Yuan}}},
  \bibinfo{journal}{\apj} \textbf{\bibinfo{volume}{523}}, \bibinfo{pages}{340}
  (\bibinfo{year}{1999}), \eprint{astro-ph/9905099}.

\bibitem[{\citenamefont{{Chakrabarti} and {Das}}(2004)}]{chakrabarti-das-2004}
\bibinfo{author}{\bibfnamefont{S.~K.} \bibnamefont{{Chakrabarti}}}
  \bibnamefont{and} \bibinfo{author}{\bibfnamefont{S.}~\bibnamefont{{Das}}},
  \bibinfo{journal}{\mnras} \textbf{\bibinfo{volume}{349}},
  \bibinfo{pages}{649} (\bibinfo{year}{2004}), \eprint{astro-ph/0402561}.

\bibitem[{\citenamefont{{Le} and {Becker}}(2004)}]{Le-Becker2004}
\bibinfo{author}{\bibfnamefont{T.}~\bibnamefont{{Le}}} \bibnamefont{and}
  \bibinfo{author}{\bibfnamefont{P.~A.} \bibnamefont{{Becker}}},
  \bibinfo{journal}{\apjl} \textbf{\bibinfo{volume}{617}}, \bibinfo{pages}{L25}
  (\bibinfo{year}{2004}), \eprint{astro-ph/0411801}.

\bibitem[{\citenamefont{{Fukumura} and {Tsuruta}}(2004)}]{Fukumura-Tsuruta2004}
\bibinfo{author}{\bibfnamefont{K.}~\bibnamefont{{Fukumura}}} \bibnamefont{and}
  \bibinfo{author}{\bibfnamefont{S.}~\bibnamefont{{Tsuruta}}},
  \bibinfo{journal}{\apj} \textbf{\bibinfo{volume}{611}}, \bibinfo{pages}{964}
  (\bibinfo{year}{2004}), \eprint{astro-ph/0405269}.

\bibitem[{\citenamefont{{Becker} et~al.}(2008)\citenamefont{{Becker}, {Das},
  and {Le}}}]{Becker-etal2008}
\bibinfo{author}{\bibfnamefont{P.~A.} \bibnamefont{{Becker}}},
  \bibinfo{author}{\bibfnamefont{S.}~\bibnamefont{{Das}}}, \bibnamefont{and}
  \bibinfo{author}{\bibfnamefont{T.}~\bibnamefont{{Le}}},
  \bibinfo{journal}{\apjl} \textbf{\bibinfo{volume}{677}}, \bibinfo{pages}{L93}
  (\bibinfo{year}{2008}), \eprint{0907.0872}.

\bibitem[{\citenamefont{{Das} et~al.}(2009)\citenamefont{{Das}, {Becker}, and
  {Le}}}]{Das-etal2009}
\bibinfo{author}{\bibfnamefont{S.}~\bibnamefont{{Das}}},
  \bibinfo{author}{\bibfnamefont{P.~A.} \bibnamefont{{Becker}}},
  \bibnamefont{and} \bibinfo{author}{\bibfnamefont{T.}~\bibnamefont{{Le}}},
  \bibinfo{journal}{\apj} \textbf{\bibinfo{volume}{702}}, \bibinfo{pages}{649}
  (\bibinfo{year}{2009}), \eprint{0907.0875}.

\bibitem[{\citenamefont{{Kumar} et~al.}(2013)\citenamefont{{Kumar}, {Singh},
  {Chattopadhyay}, and {Chakrabarti}}}]{Kumar-etal2013}
\bibinfo{author}{\bibfnamefont{R.}~\bibnamefont{{Kumar}}},
  \bibinfo{author}{\bibfnamefont{C.~B.} \bibnamefont{{Singh}}},
  \bibinfo{author}{\bibfnamefont{I.}~\bibnamefont{{Chattopadhyay}}},
  \bibnamefont{and} \bibinfo{author}{\bibfnamefont{S.~K.}
  \bibnamefont{{Chakrabarti}}}, \bibinfo{journal}{\mnras}
  \textbf{\bibinfo{volume}{436}}, \bibinfo{pages}{2864} (\bibinfo{year}{2013}),
  \eprint{1310.0144}.

\bibitem[{\citenamefont{{Sarkar} and {Das}}(2016)}]{Sarkar-Das2016}
\bibinfo{author}{\bibfnamefont{B.}~\bibnamefont{{Sarkar}}} \bibnamefont{and}
  \bibinfo{author}{\bibfnamefont{S.}~\bibnamefont{{Das}}},
  \bibinfo{journal}{\mnras} \textbf{\bibinfo{volume}{461}},
  \bibinfo{pages}{190} (\bibinfo{year}{2016}), \eprint{1606.00526}.

\bibitem[{\citenamefont{{Aktar} et~al.}(2017)\citenamefont{{Aktar}, {Das},
  {Nandi}, and {Sreehari}}}]{Aktar-etal2017}
\bibinfo{author}{\bibfnamefont{R.}~\bibnamefont{{Aktar}}},
  \bibinfo{author}{\bibfnamefont{S.}~\bibnamefont{{Das}}},
  \bibinfo{author}{\bibfnamefont{A.}~\bibnamefont{{Nandi}}}, \bibnamefont{and}
  \bibinfo{author}{\bibfnamefont{H.}~\bibnamefont{{Sreehari}}},
  \bibinfo{journal}{\mnras} \textbf{\bibinfo{volume}{471}},
  \bibinfo{pages}{4806} (\bibinfo{year}{2017}), \eprint{1707.07511}.

\bibitem[{\citenamefont{{Dihingia} et~al.}(2019)\citenamefont{{Dihingia},
  {Das}, and {Nandi}}}]{dihingia-etal-2019A}
\bibinfo{author}{\bibfnamefont{I.~K.} \bibnamefont{{Dihingia}}},
  \bibinfo{author}{\bibfnamefont{S.}~\bibnamefont{{Das}}}, \bibnamefont{and}
  \bibinfo{author}{\bibfnamefont{A.}~\bibnamefont{{Nandi}}},
  \bibinfo{journal}{Monthly Notices of the Royal Astronomical Society}
  \textbf{\bibinfo{volume}{484}}, \bibinfo{pages}{3209} (\bibinfo{year}{2019}),
  \eprint{1901.04293}.

\bibitem[{\citenamefont{{Sen} et~al.}(2022)\citenamefont{{Sen}, {Maity}, and
  {Das}}}]{Sen-etal2022}
\bibinfo{author}{\bibfnamefont{G.}~\bibnamefont{{Sen}}},
  \bibinfo{author}{\bibfnamefont{D.}~\bibnamefont{{Maity}}}, \bibnamefont{and}
  \bibinfo{author}{\bibfnamefont{S.}~\bibnamefont{{Das}}},
  \bibinfo{journal}{\jcap} \textbf{\bibinfo{volume}{2022}}, \bibinfo{eid}{048}
  (\bibinfo{year}{2022}), \eprint{2204.02110}.

\bibitem[{\citenamefont{{Singh} and {Das}}(2024)}]{Singh-Das2024}
\bibinfo{author}{\bibfnamefont{M.}~\bibnamefont{{Singh}}} \bibnamefont{and}
  \bibinfo{author}{\bibfnamefont{S.}~\bibnamefont{{Das}}},
  \bibinfo{journal}{\apss} \textbf{\bibinfo{volume}{369}}, \bibinfo{eid}{1}
  (\bibinfo{year}{2024}), \eprint{2312.16001}.

\bibitem[{\citenamefont{{Chakrabarti} and
  {Molteni}}(1993)}]{Chakrabarti-Molteni1993}
\bibinfo{author}{\bibfnamefont{S.~K.} \bibnamefont{{Chakrabarti}}}
  \bibnamefont{and}
  \bibinfo{author}{\bibfnamefont{D.}~\bibnamefont{{Molteni}}},
  \bibinfo{journal}{\apj} \textbf{\bibinfo{volume}{417}}, \bibinfo{pages}{671}
  (\bibinfo{year}{1993}), \eprint{astro-ph/9310042}.

\bibitem[{\citenamefont{{Molteni} et~al.}(1994)\citenamefont{{Molteni},
  {Lanzafame}, and {Chakrabarti}}}]{Molteni-etal1994}
\bibinfo{author}{\bibfnamefont{D.}~\bibnamefont{{Molteni}}},
  \bibinfo{author}{\bibfnamefont{G.}~\bibnamefont{{Lanzafame}}},
  \bibnamefont{and} \bibinfo{author}{\bibfnamefont{S.~K.}
  \bibnamefont{{Chakrabarti}}}, \bibinfo{journal}{\apj}
  \textbf{\bibinfo{volume}{425}}, \bibinfo{pages}{161} (\bibinfo{year}{1994}),
  \eprint{astro-ph/9310047}.

\bibitem[{\citenamefont{{Molteni} et~al.}(1996)\citenamefont{{Molteni}, {Ryu},
  and {Chakrabarti}}}]{Molteni-etal1996}
\bibinfo{author}{\bibfnamefont{D.}~\bibnamefont{{Molteni}}},
  \bibinfo{author}{\bibfnamefont{D.}~\bibnamefont{{Ryu}}}, \bibnamefont{and}
  \bibinfo{author}{\bibfnamefont{S.~K.} \bibnamefont{{Chakrabarti}}},
  \bibinfo{journal}{\apj} \textbf{\bibinfo{volume}{470}}, \bibinfo{pages}{460}
  (\bibinfo{year}{1996}), \eprint{astro-ph/9605116}.

\bibitem[{\citenamefont{{Ryu} et~al.}(1997)\citenamefont{{Ryu}, {Chakrabarti},
  and {Molteni}}}]{Ryu-etal1997}
\bibinfo{author}{\bibfnamefont{D.}~\bibnamefont{{Ryu}}},
  \bibinfo{author}{\bibfnamefont{S.~K.} \bibnamefont{{Chakrabarti}}},
  \bibnamefont{and}
  \bibinfo{author}{\bibfnamefont{D.}~\bibnamefont{{Molteni}}},
  \bibinfo{journal}{\apj} \textbf{\bibinfo{volume}{474}}, \bibinfo{pages}{378}
  (\bibinfo{year}{1997}), \eprint{astro-ph/9607051}.

\bibitem[{\citenamefont{{Lanzafame} et~al.}(1998)\citenamefont{{Lanzafame},
  {Molteni}, and {Chakrabarti}}}]{Lanzafame-etal1998}
\bibinfo{author}{\bibfnamefont{G.}~\bibnamefont{{Lanzafame}}},
  \bibinfo{author}{\bibfnamefont{D.}~\bibnamefont{{Molteni}}},
  \bibnamefont{and} \bibinfo{author}{\bibfnamefont{S.~K.}
  \bibnamefont{{Chakrabarti}}}, \bibinfo{journal}{\mnras}
  \textbf{\bibinfo{volume}{299}}, \bibinfo{pages}{799} (\bibinfo{year}{1998}),
  \eprint{astro-ph/9706248}.

\bibitem[{\citenamefont{{Das} et~al.}(2014)\citenamefont{{Das},
  {Chattopadhyay}, {Nandi}, and {Molteni}}}]{Das-etal2014}
\bibinfo{author}{\bibfnamefont{S.}~\bibnamefont{{Das}}},
  \bibinfo{author}{\bibfnamefont{I.}~\bibnamefont{{Chattopadhyay}}},
  \bibinfo{author}{\bibfnamefont{A.}~\bibnamefont{{Nandi}}}, \bibnamefont{and}
  \bibinfo{author}{\bibfnamefont{D.}~\bibnamefont{{Molteni}}},
  \bibinfo{journal}{\mnras} \textbf{\bibinfo{volume}{442}},
  \bibinfo{pages}{251} (\bibinfo{year}{2014}), \eprint{1405.4415}.

\bibitem[{\citenamefont{{Okuda} and {Das}}(2015)}]{Okuda-Das2015}
\bibinfo{author}{\bibfnamefont{T.}~\bibnamefont{{Okuda}}} \bibnamefont{and}
  \bibinfo{author}{\bibfnamefont{S.}~\bibnamefont{{Das}}},
  \bibinfo{journal}{\mnras} \textbf{\bibinfo{volume}{453}},
  \bibinfo{pages}{147} (\bibinfo{year}{2015}), \eprint{1507.04326}.

\bibitem[{\citenamefont{{Sukov{\'a}} and {Janiuk}}(2015)}]{Sukova-Janiuk2015}
\bibinfo{author}{\bibfnamefont{P.}~\bibnamefont{{Sukov{\'a}}}}
  \bibnamefont{and} \bibinfo{author}{\bibfnamefont{A.}~\bibnamefont{{Janiuk}}},
  \bibinfo{journal}{\mnras} \textbf{\bibinfo{volume}{447}},
  \bibinfo{pages}{1565} (\bibinfo{year}{2015}), \eprint{1411.7836}.

\bibitem[{\citenamefont{{Lee} et~al.}(2016)\citenamefont{{Lee},
  {Chattopadhyay}, {Kumar}, {Hyung}, and {Ryu}}}]{Lee-atal2016}
\bibinfo{author}{\bibfnamefont{S.-J.} \bibnamefont{{Lee}}},
  \bibinfo{author}{\bibfnamefont{I.}~\bibnamefont{{Chattopadhyay}}},
  \bibinfo{author}{\bibfnamefont{R.}~\bibnamefont{{Kumar}}},
  \bibinfo{author}{\bibfnamefont{S.}~\bibnamefont{{Hyung}}}, \bibnamefont{and}
  \bibinfo{author}{\bibfnamefont{D.}~\bibnamefont{{Ryu}}},
  \bibinfo{journal}{\apj} \textbf{\bibinfo{volume}{831}}, \bibinfo{eid}{33}
  (\bibinfo{year}{2016}), \eprint{1608.03997}.

\bibitem[{\citenamefont{{Sukov{\'a}} et~al.}(2017)\citenamefont{{Sukov{\'a}},
  {Charzy{\'n}ski}, and {Janiuk}}}]{Sukova-etal2017}
\bibinfo{author}{\bibfnamefont{P.}~\bibnamefont{{Sukov{\'a}}}},
  \bibinfo{author}{\bibfnamefont{S.}~\bibnamefont{{Charzy{\'n}ski}}},
  \bibnamefont{and} \bibinfo{author}{\bibfnamefont{A.}~\bibnamefont{{Janiuk}}},
  \bibinfo{journal}{\mnras} \textbf{\bibinfo{volume}{472}},
  \bibinfo{pages}{4327} (\bibinfo{year}{2017}), \eprint{1709.01824}.

\bibitem[{\citenamefont{{Kim} et~al.}(2019)\citenamefont{{Kim}, {Garain},
  {Chakrabarti}, and {Balsara}}}]{Kim-etal2019}
\bibinfo{author}{\bibfnamefont{J.}~\bibnamefont{{Kim}}},
  \bibinfo{author}{\bibfnamefont{S.~K.} \bibnamefont{{Garain}}},
  \bibinfo{author}{\bibfnamefont{S.~K.} \bibnamefont{{Chakrabarti}}},
  \bibnamefont{and} \bibinfo{author}{\bibfnamefont{D.~S.}
  \bibnamefont{{Balsara}}}, \bibinfo{journal}{\mnras}
  \textbf{\bibinfo{volume}{482}}, \bibinfo{pages}{3636} (\bibinfo{year}{2019}),
  \eprint{1810.12469}.

\bibitem[{\citenamefont{{Okuda} et~al.}(2019)\citenamefont{{Okuda}, {Singh},
  {Das}, {Aktar}, {Nandi}, and {Dal Pino}}}]{Okuda-etal2019}
\bibinfo{author}{\bibfnamefont{T.}~\bibnamefont{{Okuda}}},
  \bibinfo{author}{\bibfnamefont{C.~B.} \bibnamefont{{Singh}}},
  \bibinfo{author}{\bibfnamefont{S.}~\bibnamefont{{Das}}},
  \bibinfo{author}{\bibfnamefont{R.}~\bibnamefont{{Aktar}}},
  \bibinfo{author}{\bibfnamefont{A.}~\bibnamefont{{Nandi}}}, \bibnamefont{and}
  \bibinfo{author}{\bibfnamefont{E.~M. d.~G.} \bibnamefont{{Dal Pino}}},
  \bibinfo{journal}{\pasj} \textbf{\bibinfo{volume}{71}}, \bibinfo{eid}{49}
  (\bibinfo{year}{2019}), \eprint{1902.02933}.

\bibitem[{\citenamefont{{Palit} et~al.}(2019)\citenamefont{{Palit}, {Janiuk},
  and {Sukova}}}]{Palit-etal2019}
\bibinfo{author}{\bibfnamefont{I.}~\bibnamefont{{Palit}}},
  \bibinfo{author}{\bibfnamefont{A.}~\bibnamefont{{Janiuk}}}, \bibnamefont{and}
  \bibinfo{author}{\bibfnamefont{P.}~\bibnamefont{{Sukova}}},
  \bibinfo{journal}{\mnras} \textbf{\bibinfo{volume}{487}},
  \bibinfo{pages}{755} (\bibinfo{year}{2019}), \eprint{1905.02289}.

\bibitem[{\citenamefont{{Debnath} et~al.}(2024)\citenamefont{{Debnath},
  {Chattopadhyay}, and {Joshi}}}]{Debnath-etal2024}
\bibinfo{author}{\bibfnamefont{S.}~\bibnamefont{{Debnath}}},
  \bibinfo{author}{\bibfnamefont{I.}~\bibnamefont{{Chattopadhyay}}},
  \bibnamefont{and} \bibinfo{author}{\bibfnamefont{R.~K.}
  \bibnamefont{{Joshi}}}, \bibinfo{journal}{\mnras}
  \textbf{\bibinfo{volume}{528}}, \bibinfo{pages}{3964} (\bibinfo{year}{2024}),
  \eprint{2401.07786}.

\bibitem[{\citenamefont{{TianLe-Zhao} et~al.}(2024)\citenamefont{{TianLe-Zhao},
  {XiaoFeng-Li}, {ZeYuan-Tang}, and {Kumar}}}]{TianLe-Zhao-etal2024}
\bibinfo{author}{\bibnamefont{{TianLe-Zhao}}},
  \bibinfo{author}{\bibnamefont{{XiaoFeng-Li}}},
  \bibinfo{author}{\bibnamefont{{ZeYuan-Tang}}}, \bibnamefont{and}
  \bibinfo{author}{\bibfnamefont{R.}~\bibnamefont{{Kumar}}},
  \bibinfo{journal}{arXiv e-prints} \bibinfo{eid}{arXiv:2407.01859}
  (\bibinfo{year}{2024}), \eprint{2407.01859}.

\bibitem[{\citenamefont{{Chakrabarti} and
  {Titarchuk}}(1995)}]{chakrabarti-titarchuk-1995}
\bibinfo{author}{\bibfnamefont{S.}~\bibnamefont{{Chakrabarti}}}
  \bibnamefont{and} \bibinfo{author}{\bibfnamefont{L.~G.}
  \bibnamefont{{Titarchuk}}}, \bibinfo{journal}{\apj}
  \textbf{\bibinfo{volume}{455}}, \bibinfo{pages}{623} (\bibinfo{year}{1995}),
  \eprint{astro-ph/9510005}.

\bibitem[{\citenamefont{{Nandi} et~al.}(2012)\citenamefont{{Nandi}, {Debnath},
  {Mandal}, and {Chakrabarti}}}]{nandi-etal-2012}
\bibinfo{author}{\bibfnamefont{A.}~\bibnamefont{{Nandi}}},
  \bibinfo{author}{\bibfnamefont{D.}~\bibnamefont{{Debnath}}},
  \bibinfo{author}{\bibfnamefont{S.}~\bibnamefont{{Mandal}}}, \bibnamefont{and}
  \bibinfo{author}{\bibfnamefont{S.~K.} \bibnamefont{{Chakrabarti}}},
  \bibinfo{journal}{\aap} \textbf{\bibinfo{volume}{542}}, \bibinfo{eid}{A56}
  (\bibinfo{year}{2012}), \eprint{1204.5044}.

\bibitem[{\citenamefont{{Iyer} et~al.}(2015)\citenamefont{{Iyer}, {Nandi}, and
  {Mandal}}}]{Iyer-etal2015}
\bibinfo{author}{\bibfnamefont{N.}~\bibnamefont{{Iyer}}},
  \bibinfo{author}{\bibfnamefont{A.}~\bibnamefont{{Nandi}}}, \bibnamefont{and}
  \bibinfo{author}{\bibfnamefont{S.}~\bibnamefont{{Mandal}}},
  \bibinfo{journal}{\apj} \textbf{\bibinfo{volume}{807}}, \bibinfo{eid}{108}
  (\bibinfo{year}{2015}), \eprint{1505.02529}.

\bibitem[{\citenamefont{{Nandi} et~al.}(2018)\citenamefont{{Nandi}, {Mandal},
  {Sreehari}, {Radhika}, {Das}, {Chattopadhyay}, {Iyer}, {Agrawal}, and
  {Aktar}}}]{Nandi-etal2018}
\bibinfo{author}{\bibfnamefont{A.}~\bibnamefont{{Nandi}}},
  \bibinfo{author}{\bibfnamefont{S.}~\bibnamefont{{Mandal}}},
  \bibinfo{author}{\bibfnamefont{H.}~\bibnamefont{{Sreehari}}},
  \bibinfo{author}{\bibfnamefont{D.}~\bibnamefont{{Radhika}}},
  \bibinfo{author}{\bibfnamefont{S.}~\bibnamefont{{Das}}},
  \bibinfo{author}{\bibfnamefont{I.}~\bibnamefont{{Chattopadhyay}}},
  \bibinfo{author}{\bibfnamefont{N.}~\bibnamefont{{Iyer}}},
  \bibinfo{author}{\bibfnamefont{V.~K.} \bibnamefont{{Agrawal}}},
  \bibnamefont{and} \bibinfo{author}{\bibfnamefont{R.}~\bibnamefont{{Aktar}}},
  \bibinfo{journal}{\apss} \textbf{\bibinfo{volume}{363}}, \bibinfo{eid}{90}
  (\bibinfo{year}{2018}), \eprint{1803.08638}.

\bibitem[{\citenamefont{{Das} et~al.}(2021{\natexlab{a}})\citenamefont{{Das},
  {Nandi}, {Agrawal}, {Dihingia}, and {Majumder}}}]{Das-etal2021}
\bibinfo{author}{\bibfnamefont{S.}~\bibnamefont{{Das}}},
  \bibinfo{author}{\bibfnamefont{A.}~\bibnamefont{{Nandi}}},
  \bibinfo{author}{\bibfnamefont{V.~K.} \bibnamefont{{Agrawal}}},
  \bibinfo{author}{\bibfnamefont{I.~K.} \bibnamefont{{Dihingia}}},
  \bibnamefont{and}
  \bibinfo{author}{\bibfnamefont{S.}~\bibnamefont{{Majumder}}},
  \bibinfo{journal}{\mnras} \textbf{\bibinfo{volume}{507}},
  \bibinfo{pages}{2777} (\bibinfo{year}{2021}{\natexlab{a}}),
  \eprint{2108.02973}.

\bibitem[{\citenamefont{{Majumder} et~al.}(2023)\citenamefont{{Majumder},
  {Das}, {Agrawal}, and {Nandi}}}]{Majumder-etal2023}
\bibinfo{author}{\bibfnamefont{S.}~\bibnamefont{{Majumder}}},
  \bibinfo{author}{\bibfnamefont{S.}~\bibnamefont{{Das}}},
  \bibinfo{author}{\bibfnamefont{V.~K.} \bibnamefont{{Agrawal}}},
  \bibnamefont{and} \bibinfo{author}{\bibfnamefont{A.}~\bibnamefont{{Nandi}}},
  \bibinfo{journal}{\mnras} \textbf{\bibinfo{volume}{526}},
  \bibinfo{pages}{2086} (\bibinfo{year}{2023}), \eprint{2309.11182}.

\bibitem[{\citenamefont{{Event Horizon Telescope Collaboration}
  et~al.}(2021)\citenamefont{{Event Horizon Telescope Collaboration},
  {Akiyama}, {Algaba}, {Alberdi}, {Alef}, {Anantua}, {Asada}, {Azulay},
  {Baczko}, {Ball} et~al.}}]{EHT-M87-2021}
\bibinfo{author}{\bibnamefont{{Event Horizon Telescope Collaboration}}},
  \bibinfo{author}{\bibfnamefont{K.}~\bibnamefont{{Akiyama}}},
  \bibinfo{author}{\bibfnamefont{J.~C.} \bibnamefont{{Algaba}}},
  \bibinfo{author}{\bibfnamefont{A.}~\bibnamefont{{Alberdi}}},
  \bibinfo{author}{\bibfnamefont{W.}~\bibnamefont{{Alef}}},
  \bibinfo{author}{\bibfnamefont{R.}~\bibnamefont{{Anantua}}},
  \bibinfo{author}{\bibfnamefont{K.}~\bibnamefont{{Asada}}},
  \bibinfo{author}{\bibfnamefont{R.}~\bibnamefont{{Azulay}}},
  \bibinfo{author}{\bibfnamefont{A.-K.} \bibnamefont{{Baczko}}},
  \bibinfo{author}{\bibfnamefont{D.}~\bibnamefont{{Ball}}},
  \bibnamefont{et~al.}, \bibinfo{journal}{\apjl}
  \textbf{\bibinfo{volume}{910}}, \bibinfo{eid}{L13} (\bibinfo{year}{2021}),
  \eprint{2105.01173}.

\bibitem[{\citenamefont{{Event Horizon Telescope Collaboration}
  et~al.}(2022)\citenamefont{{Event Horizon Telescope Collaboration},
  {Akiyama}, {Alberdi}, {Alef}, {Algaba}, {Anantua}, {Asada}, {Azulay}, {Bach},
  {Baczko} et~al.}}]{EHT-SGRA*-2022}
\bibinfo{author}{\bibnamefont{{Event Horizon Telescope Collaboration}}},
  \bibinfo{author}{\bibfnamefont{K.}~\bibnamefont{{Akiyama}}},
  \bibinfo{author}{\bibfnamefont{A.}~\bibnamefont{{Alberdi}}},
  \bibinfo{author}{\bibfnamefont{W.}~\bibnamefont{{Alef}}},
  \bibinfo{author}{\bibfnamefont{J.~C.} \bibnamefont{{Algaba}}},
  \bibinfo{author}{\bibfnamefont{R.}~\bibnamefont{{Anantua}}},
  \bibinfo{author}{\bibfnamefont{K.}~\bibnamefont{{Asada}}},
  \bibinfo{author}{\bibfnamefont{R.}~\bibnamefont{{Azulay}}},
  \bibinfo{author}{\bibfnamefont{U.}~\bibnamefont{{Bach}}},
  \bibinfo{author}{\bibfnamefont{A.-K.} \bibnamefont{{Baczko}}},
  \bibnamefont{et~al.}, \bibinfo{journal}{\apjl}
  \textbf{\bibinfo{volume}{930}}, \bibinfo{eid}{L16} (\bibinfo{year}{2022}).

\bibitem[{\citenamefont{{Quataert}}(2008)}]{Quataert2008}
\bibinfo{author}{\bibfnamefont{E.}~\bibnamefont{{Quataert}}},
  \bibinfo{journal}{\apj} \textbf{\bibinfo{volume}{673}}, \bibinfo{pages}{758}
  (\bibinfo{year}{2008}), \eprint{0710.5521}.

\bibitem[{\citenamefont{{Dihingia} et~al.}(2018)\citenamefont{{Dihingia},
  {Das}, {Maity}, and {Chakrabarti}}}]{dihingia-etal-2018}
\bibinfo{author}{\bibfnamefont{I.~K.} \bibnamefont{{Dihingia}}},
  \bibinfo{author}{\bibfnamefont{S.}~\bibnamefont{{Das}}},
  \bibinfo{author}{\bibfnamefont{D.}~\bibnamefont{{Maity}}}, \bibnamefont{and}
  \bibinfo{author}{\bibfnamefont{S.}~\bibnamefont{{Chakrabarti}}},
  \bibinfo{journal}{\prd} \textbf{\bibinfo{volume}{98}}, \bibinfo{eid}{083004}
  (\bibinfo{year}{2018}), \eprint{1806.08481}.

\bibitem[{\citenamefont{{Riffert} and {Herold}}(1995)}]{Riffert-Herold1995}
\bibinfo{author}{\bibfnamefont{H.}~\bibnamefont{{Riffert}}} \bibnamefont{and}
  \bibinfo{author}{\bibfnamefont{H.}~\bibnamefont{{Herold}}},
  \bibinfo{journal}{\apj} \textbf{\bibinfo{volume}{450}}, \bibinfo{pages}{508}
  (\bibinfo{year}{1995}).

\bibitem[{\citenamefont{{Peitz} and {Appl}}(1997)}]{Peitz-Appl1997}
\bibinfo{author}{\bibfnamefont{J.}~\bibnamefont{{Peitz}}} \bibnamefont{and}
  \bibinfo{author}{\bibfnamefont{S.}~\bibnamefont{{Appl}}},
  \bibinfo{journal}{\mnras} \textbf{\bibinfo{volume}{286}},
  \bibinfo{pages}{681} (\bibinfo{year}{1997}), \eprint{astro-ph/9612205}.

\bibitem[{\citenamefont{{Cowie} and {McKee}}(1977)}]{cowie-mcKee-1977}
\bibinfo{author}{\bibfnamefont{L.~L.} \bibnamefont{{Cowie}}} \bibnamefont{and}
  \bibinfo{author}{\bibfnamefont{C.~F.} \bibnamefont{{McKee}}},
  \bibinfo{journal}{\apj} \textbf{\bibinfo{volume}{211}}, \bibinfo{pages}{135}
  (\bibinfo{year}{1977}).

\bibitem[{\citenamefont{{Chattopadhyay} and
  {Ryu}}(2009)}]{chattopadhyay-ryu-2009}
\bibinfo{author}{\bibfnamefont{I.}~\bibnamefont{{Chattopadhyay}}}
  \bibnamefont{and} \bibinfo{author}{\bibfnamefont{D.}~\bibnamefont{{Ryu}}},
  \bibinfo{journal}{\apj} \textbf{\bibinfo{volume}{694}}, \bibinfo{pages}{492}
  (\bibinfo{year}{2009}), \eprint{0812.2607}.

\bibitem[{\citenamefont{{Das} et~al.}(2021{\natexlab{b}})\citenamefont{{Das},
  {Nandi}, {Agrawal}, {Dihingia}, and {Majumder}}}]{das-etal-2021}
\bibinfo{author}{\bibfnamefont{S.}~\bibnamefont{{Das}}},
  \bibinfo{author}{\bibfnamefont{A.}~\bibnamefont{{Nandi}}},
  \bibinfo{author}{\bibfnamefont{V.~K.} \bibnamefont{{Agrawal}}},
  \bibinfo{author}{\bibfnamefont{I.~K.} \bibnamefont{{Dihingia}}},
  \bibnamefont{and}
  \bibinfo{author}{\bibfnamefont{S.}~\bibnamefont{{Majumder}}},
  \bibinfo{journal}{\mnras} \textbf{\bibinfo{volume}{507}},
  \bibinfo{pages}{2777} (\bibinfo{year}{2021}{\natexlab{b}}),
  \eprint{2108.02973}.

\bibitem[{\citenamefont{{Fukue}}(2019)}]{Fukue2019}
\bibinfo{author}{\bibfnamefont{J.}~\bibnamefont{{Fukue}}},
  \bibinfo{journal}{\mnras} \textbf{\bibinfo{volume}{483}},
  \bibinfo{pages}{2538} (\bibinfo{year}{2019}).

\bibitem[{\citenamefont{{Holzer} and {Axford}}(1970)}]{Holzer-Axford1970}
\bibinfo{author}{\bibfnamefont{T.~E.} \bibnamefont{{Holzer}}} \bibnamefont{and}
  \bibinfo{author}{\bibfnamefont{W.~I.} \bibnamefont{{Axford}}},
  \bibinfo{journal}{\araa} \textbf{\bibinfo{volume}{8}}, \bibinfo{pages}{31}
  (\bibinfo{year}{1970}).

\bibitem[{\citenamefont{{Landau} and {Lifshitz}}(1959)}]{Landau-Lifshitz1959}
\bibinfo{author}{\bibfnamefont{L.~D.} \bibnamefont{{Landau}}} \bibnamefont{and}
  \bibinfo{author}{\bibfnamefont{E.~M.} \bibnamefont{{Lifshitz}}},
  \emph{\bibinfo{title}{{Fluid mechanics}}} (\bibinfo{publisher}{New York:
  Pergamon}, \bibinfo{year}{1959}).

\bibitem[{\citenamefont{{Chakrabarti}}(1990)}]{Chakrabarti-1990}
\bibinfo{author}{\bibfnamefont{S.~K.} \bibnamefont{{Chakrabarti}}},
  \emph{\bibinfo{title}{{Theory of Transonic Astrophysical Flows}}}
  (\bibinfo{publisher}{World Scientific Publishing}, \bibinfo{year}{1990}).

\end{thebibliography}

\section*{Appendix: Derivation of wind equation}

After some algebra, the radial momentum, azimuthal momentum and entropy generation equations read as, 
$$
R_{0}+ R_{\upsilon}\frac{d \upsilon}{d r} + R_{\Theta}\frac{d \Theta}{d r} + R_{\lambda}\frac{d \lambda}{d r} = 0 ,
$$
$$
L_{0} + L_{\upsilon}\frac{d \upsilon}{d r} + L_{\Theta}\frac{d \Theta}{d r} + L_{\lambda}\frac{d \lambda}{d r} = 0 ,
$$
$$
E_{0} + E_{\upsilon}\frac{d \upsilon}{d r} + E_{\Theta}\frac{d \Theta}{d r} + E_{\lambda}\frac{d \lambda}{d r} = 0.
$$
After further simplification, we have,
\begin{equation*}
\frac{d\upsilon}{d r} = \frac{\mathcal{N}}{\mathcal{D}},
\end{equation*}
\begin{equation*}
\frac{d\Theta}{d r} = \Theta_{1}+\Theta_{2}\frac{d\upsilon}{d r}, 
\end{equation*}
\begin{equation*}
\frac{d\lambda}{d r} = \lambda_{1}+\lambda_{2}\frac{d\upsilon}{d r},
\end{equation*}
where
\begin{equation*}
   \begin{aligned}
\mathcal{N} ={} & E_{\lambda}\left(-R_{\Theta} L_{0}+ R_{0} L_{\Theta}\right) + E_{\Theta}\left(R_{\lambda} L_{0}- R_{0} L_{\lambda}\right) \\ &+ E_{0}\left(-R_{\lambda} L_{\Theta}+ R_{\Theta} L_{\lambda}\right) ,
\end{aligned}
\end{equation*}

\begin{equation*}
   \begin{aligned}
\mathcal{D} ={} &  E_{\lambda}\left(R_{\Theta} L_{\upsilon} - R_{\upsilon} L_{\Theta}\right) + E_{\Theta} \left(-R_{\lambda} L_{\upsilon}+ R_{\upsilon} L_{\lambda}\right) \\ &+ E_{\upsilon}\left(R_{\lambda} L_{\Theta} -  R_{\Theta} L_{\lambda}\right).
\end{aligned}
\end{equation*}

The expressions of the coefficients used in the above equations are given by,
$$
\Theta_{1} = \frac{\Theta_{11}}{\Theta_{33}}, \thickspace \thickspace \Theta_{2} = \frac{\Theta_{22}}{\Theta_{33}} ,\thickspace\thickspace \lambda_{1} = \frac{\lambda_{11}}{\Theta_{33}}, \thickspace\thickspace \lambda_{2} = \frac{\lambda_{22}}{\Theta_{33}} ,
$$

\begin{equation*}
\begin{aligned}
\Theta_{11} = &	E_{\lambda} L_{0}- E_{0} L_{\lambda},\thickspace\thickspace \Theta_{22} = E_{\lambda} L_{\upsilon} - E_{\upsilon} L_{\lambda},\thickspace \\ \Theta_{33} = & - E_{\lambda} L_{\Theta} + E_{\Theta} L_{\lambda},
\end{aligned}
\end{equation*}

$$
\lambda_{11} = 	-E_{\Theta} L_{0}+ E_{0} L_{\Theta},\thickspace\thickspace \lambda_{22} = -E_{\Theta} L_{\upsilon} + E_{\upsilon} L_{\Theta},
$$

\begin{equation*}
\begin{aligned}
R_{0} = & \frac{d \Phi_{e}^{\rm{eff}}}{d r} - \frac{3\Theta}{r \tau h} + \frac{F_{3}\Theta}{\tau F_{2} h} - \frac{\Theta}{\tau \Delta h}\frac{d \Delta}{d r}, \\ R_{\Theta} = &  \frac{1}{\tau h},\thickspace  R_{\lambda} = \frac{F_{4} \Theta}{\tau F_{2} h },\thickspace R_{\upsilon} = \upsilon -\frac{2\Theta}{\tau \upsilon h},
\end{aligned}
\end{equation*}

\begin{equation*}
	\begin{aligned}
E_{0} = & \frac{5 \Phi_{\rm{s}}}{2} \left(\frac{2 \Theta}{\tau}\right)^{3/2} \left(\frac{1}{r} - \frac{F_{3}}{F_{2}} + \frac{1}{\Delta} \frac{d \Delta}{d r} \right)  -r \alpha \upsilon^{2}\omega_{1} \\ &- \frac{2 r \alpha\Theta \omega_{1}}{\tau} + \frac{3 \upsilon \Theta}{r \tau} - \frac{F_{3} \upsilon \Theta}{\tau F_{2}} + \frac{\upsilon \Theta}{\tau \Delta} \frac{d \Delta}{d r},
\end{aligned}
\end{equation*}

$$
E_{\Theta} = -\frac{5 \Phi_{\rm{s}}}{\Theta} \left(\frac{2 \Theta}{\tau}\right)^{3/2}  + \frac{\left(1+ 2 N\right) \upsilon}{\tau}
$$

\begin{equation*}
	\begin{aligned}
E_{\lambda} = & \frac{5 \Phi_{\rm{s}} F_{4}}{2 F_{2}} \left(\frac{2 \Theta}{\tau}\right)^{3/2} - \frac{F_{4} \upsilon \Theta}{\tau F_{2}} - r \upsilon^2 \alpha \omega_{2} - \frac{2 r \alpha \Theta \omega_{2}}{\tau}, \\ E_{\upsilon} = & \frac{5 \Phi_{\rm{s}}}{\upsilon} \left(\frac{2 \Theta}{\tau}\right)^{3/2} + \frac{2 \Theta}{\tau},
\end{aligned}
\end{equation*}

\begin{equation*}
	\begin{aligned}
L_{0}= & -2\alpha \upsilon^{2} - \frac{4\alpha\Theta}{\tau} + \frac{r \alpha \upsilon^{2}}{2 \Delta}\frac{d \Delta}{d r} + \frac{r \alpha\Theta}{\tau \Delta} \frac{d \Delta}{d r},
L_{\Theta} = -\frac{2 r \alpha}{\tau},\\ L_{\lambda} = & \upsilon,\thickspace L_{\upsilon}  = -r \alpha \upsilon + \frac{2 r \alpha \Theta}{\tau \upsilon},
\end{aligned}
\end{equation*}

\begin{equation*}
	\begin{aligned}
	F_{1} =& \frac{\left((r^{2}+a_{\textrm{k}}^{2})^{2} + 2 \Delta a_{\textrm{k}}^2\right)}{\left((r^{2}+a_{\textrm{k}}^{2})^{2} - 2 \Delta a_{\textrm{k}}^2\right)}, F_{2} = \frac{1}{(1-\lambda\Omega)} F_{1}, \\
	F_{3} = & \frac{F_{1} \lambda \omega_{1}}{(1-\lambda\Omega)^{2}} +  \frac{1}{1-\lambda\Omega} \frac{d F_{1}}{d r},\thickspace F_{4} = \frac{F_{1} \Omega}{(1-\lambda\Omega)^{2}} + \frac{F_{1} \lambda \omega_{2}}{(1-\lambda\Omega)^{2}}, 
\end{aligned}
\end{equation*}

$$
 \frac{d 	F_{2}}{dr} = F_{3} + F_{4}\frac{d \lambda}{d r},\thickspace \frac{d \Omega}{d r} = \omega_{1}+\omega_{2}\frac{d \lambda}{d r},
$$

\begin{equation*}
	\begin{aligned}
\omega_{1} = & -\frac{2\left(a_{\textrm{k}}^{3} + 3 a_{\textrm{k}} r^{2}+ \lambda(a_{\textrm{k}} \lambda - 2 a_{\textrm{k}}^{2} + r^{2}(r-3))\right)}{\left(r^{3} + a_{\textrm{k}}^2 (r+2) - 2 a_{\textrm{k}} \lambda\right)^{2}}, \\
\omega_{2} = & \frac{r^{2}\left(a_{\textrm{k}}^2 + r (r - 2)\right)}{\left(r^{3} + a_{\textrm{k}}^2 (r+2) - 2 a_{\textrm{k}} \lambda\right)^{2}}.
\end{aligned}
\end{equation*}

\end{document}